\documentclass[nofootinbib,amsmath,amssymb,aps,pre,showkeys,onecolumn,eqsecnum,notitlepage]{revtex4-1}
\usepackage{graphicx}  
\usepackage{bm}        

\begin{document}

\title{Statistical symmetry restoration in fully developed turbulence: Renormalization group analysis of two models}

\author{N.~V.~Antonov}
\email{n.antonov@spbu.ru}
\author{N.~M.~Gulitskiy}
\email{n.gulitskiy@spbu.ru}
\author{M.~M.~Kostenko}
\email{m.m.kostenko@mail.ru}
\author{A.~V.~Malyshev}
\email{a.v.malyshev@spbu.ru}

\affiliation{Department of Physics, St. Petersburg State University, 7/9 Universitetskaya nab., St. Petersburg 199034, Russia}

\begin{abstract}
In this paper we consider the model of incompressible fluid described by the stochastic 
Navier-Stokes equation with finite correlation time of a random force.
Inertial-range asymptotic behavior of fully developed turbulence is studied 
by means of the field theoretic renormalization group within the one-loop approximation.
It is corroborated that regardless of the values of model parameters and initial data, 
the inertial-range behavior of the model is described by limiting case of
vanishing correlation time. 
It indicates that the 
Galilean symmetry of the model violated by the ``colored'' random force
is restored in the inertial range.
This regime corresponds
to the only nontrivial fixed point of the renormalization group equation.
The stability of this point depends on the relation between the exponents in the energy spectrum ${\cal E} \propto k^{1-y}$ and the dispersion law $\omega \propto k^{2-\eta}$.
The second analyzed 
problem is the passive advection of a scalar field by this velocity ensemble. 
Correlation functions of the scalar field exhibit anomalous scaling behavior in the inertial-convective range.
We demonstrate that in accordance with Kolmogorov's hypothesis of the local symmetry restoration, 
the main contribution to the operator product expansion is given by the isotropic operator, while 
anisotropic terms should be considered only as corrections.
\end{abstract}

\keywords{turbulence, renormalization group, passive scalar advection, scaling}


\maketitle


\section{Introduction}

The very important phenomenological concepts in the theory of fluid turbulence is the concept of 
the statistical symmetry restoration~\cite{Legacy,LegacyD}. The Navier-Stokes (NS) equation describing fluid dynamics 
possesses a number of symmetries: translational and rotational covariance and covariance with respect
to Galilean transformation. Some of these symmetries are violated inevitably by the experimental setup
or, speaking mathematically, by initial and boundary conditions. Moreover, some of the remaining 
symmetries can be broken spontaneously as the Reynolds number $\mathrm{Re}$ increases; see
Refs.~\cite{Legacy,LegacyD} for detailed discussions and examples. 

However, for fully developed turbulence ($\mathrm{Re}\gg1$) these symmetries can be restored in the statistical
sense, that is, for various correlation or structure functions and in a proper range of scales (inertial interval).
This concept dates back to Kolmogorov's idea of locally isotropic turbulence~\cite{Monin,K41a,K41b,K41c} and
is intuitively explained by the idea of the Richardson cascade~\cite{Richardson}. According to 
Richardson the energy fed into the system at very large scales 
is transferred downscales through numerous fractions of turbulent eddies. Thus, eventually, at small scales 
the system ``forgets'' about details of the energy pumping;
see Refs.~\cite{E84,S97,P98,P99,P00,B96} and review paper~\cite{P05} for some examples of such behavior.

In spite of their great value, phenomenological theories might appear
to be inaccurate or even incorrect. Nowadays
it is widely accepted that
due to phenomenon of intermittency, correlation functions of developed turbulence depend on the integral 
(external) scale $L$, which is in disagreement with the classical Kolmogorov K41 theory. 
In particular, the equal-time structure functions of the velocity field or those of advected scalar
field are described by an infinite set of independent anomalous exponents; see Refs.~\cite{Legacy,LegacyD}
for detailed discussion and references.

Thus, it is desirable to test predictions of phenomenological theories on the basis of certain specific
models and by means of some analytical tools. An example of such simplified but very fruitful model 
is provided by Kraichnan's rapid-change model, where the existence of anomalous scaling was
demonstrated by several approaches, and the corresponding exponents were calculated both numerically
and analytically, within controlled approximations and regular perturbation expansions; see 
Ref.~\cite{FGV} for a review and references.

The effective approach to the discussed problem is the field theoretic
renormalization group (RG), see the monographs~\cite{Zinn,Vasiliev,Tauber} and review paper~\cite{HHL}.
Kraichnan's model allows one to construct a controlled
expansion for the anomalous exponents, which 
is similar to the famous epsilon-expansion in the theory of critical
state~\cite{AAV,P96}. The practical calculations were performed up to the third (three-loop) order~\cite{cube}. 
What is more important, the model can be generalized to the more realistic cases: finite correlation time 
and non-Gaussianity of the advecting velocity field~\cite{FinTime,AKens,JphysA,FinTimeEta,Lanotte2-mod2,AGM}, 
strong anisotropy~\cite{R2,R2M,Arp,VectorN}, compressibility~\cite{Tomas,Marian2C,AK14,AK15}, helicity~\cite{HZ,MMZ,MMZR,R1}, and so on~\cite{Marian,AntGul2012,Marian2,AK2}. Admittedly, 
in application to the NS equation itself the RG method has so far restricted success~\cite{turbo}.

In this paper, we study two analytic examples of statistical symmetry restoration in fully 
developed turbulence, based on the stochastic NS equation. The fluid is assumed to be incompressible. As
it is standard for the RG approach, we consider the NS equation subjected to an 
external stirring random 
force with the prescribed Gaussian statistics. In majority of studies, the 
random force is $\delta$ correlated
in time (``white noise''), which is dictated by the Galilean symmetry. Here, we do not assume this
symmetry in advance and choose the force with finite correlation time (``colored noise''). 
This gives much more freedom for the form of the correlation function. For definiteness, we take 
the random force as the variant of the Ornstein-Uhlenbeck process~\cite{OU1,OU2}. Such statistics 
were used earlier in Ref.~\cite{NonG} for generation of the velocity field itself. 
In our case, however, the velocity field is not described within a certain
statistical ensemble but is
determined by the real nonlinear NS equation. 
Various aspects of finite correlation time of a random noise in stochastic dynamics were
discussed earlier, e.g., in Refs.~\cite{Carati,CaratiV,CaratiK,Astro1,Astro2}.

The model can be reformulated as a multiplicatively renormalizable quantum field theory. It
is well known that in such case
the possible large-scale, long-distance asymptotic regimes 
are associated with infrared (IR) attractive fixed points of the RG equations. We perform the
leading-order (one-loop) calculation and show that the only nontrivial IR attractive fixed
point corresponds to the $\delta$ correlated force. From a physical point of view it means that
the Galilean symmetry is restored immediately in the IR range, which is in 
accordance with the general concept.

The second problem is the passive advection of a scalar quantity (temperature, density of a pollutant, etc.) 
in a turbulent fluid. The latter is described by the previously studied stochastic NS field. In view of
the aforementioned result, the random force is taken to be $\delta$ correlated in time. The scalar field 
is governed by the standard advection-diffusion equation with a random force. The latter maintains the
steady state and is a source of a large-scale anisotropy. The corresponding field theory is renormalizable 
and possesses the only IR attractive fixed point. 
It is well known that correlation functions of the scalar field demonstrate anomalous scaling, so that 
the K41 theory does not hold; see Ref.~\cite{Kim} for two-loop calculation in the isotropic case. In the
RG approach, the anomalous exponents are identified with scaling dimensions of certain composite 
fields (``composite operators'' in the quantum-field terminology). For anisotropic case, it is natural 
to expand the structure functions in spherical harmonics $Y_{lm}$, where $l$ can be viewed as a degree
of anisotropy of the given contribution. In the present model, a special anomalous exponent can be assigned 
to any contribution. 

In this paper we restrict ourselves to the pair correlation function and calculate anomalous exponents in
the one-loop approximation for all $l$. It turns out, that these exponents exhibit a kind of hierarchy:
they increase monotonically with $l$. As a result, turbulence becomes less and less anisotropic in the 
depth of the inertial range, and the leading-order term is given by the isotropic contribution. 
Similar effect was observed earlier in the models of scalar and vector advection by ``synthetic'' Gaussian 
velocity fields, see Refs.~\cite{LM,ALM,FinTime,AHMMG} and literature cited therein.

The paper is organized as follows.
In Sec.~\ref{sec:Model} a detailed description of the stochastic Navier-Stokes equation for an
incompressible fluid is given.
Sec.~\ref{sec:QFT} is devoted to the field theoretic formulation of the model
and the corresponding diagrammatic technique. In particular, possible types of divergent Green's functions
are discussed.
In Sec.~\ref{RenModel} the renormalizability of the model is proven.
One-loop explicit expressions for the renormalization constants are presented and 
RG functions (anomalous dimensions and $\beta$ functions) are derived and analyzed. 
In Sec.~\ref{sec:RC} the IR asymptotic behavior, obtained by solving the RG equations, is discussed. It 
is shown that, depending on two exponents $y$ and $\eta$ 
that describe the energy spectrum and dispersion law of the velocity field,
the RG equations exhibit two nontrivial fixed points, but only one of them is stable in
the IR region. 
This means that the Galilean symmetry of the model violated by the colored random force
is restored in the inertial range.
In Sec.~\ref{Sec:CD} the corresponding scaling dimensions of the fields are presented.

In Sec.~\ref{sec:adv} an advection of a passive scalar field by the incompressible velocity field which 
obeys the NS equation is analyzed. 
The field theoretic formulation of the full model is presented. 
The existence of a scaling regime in the IR range is established. 
In Sec.~\ref{sec:ope} the operator product expansion for the pair correlation function is carried out. 
Sec.~\ref{sec:final} is devoted to the renormalization of composite operators. An inertial-range behavior
of the correlation functions is studied. 
It is shown that the leading terms of the inertial-range behavior are determined by the contributions which 
correspond to the isotropic terms. 

Sec.~\ref{sec:concl} is reserved for conclusions.
The main one is that the Galilean symmetry and isotropy, broken by introducing the external stochastic force 
with finite correlation time and by distinguished direction, are restored in the statistical sense (for
measurable quantities) in the inertial range of fully developed turbulence.

Appendix~\ref{calc} contains detailed calculations of the diagrams, needed 
to perform multiplicative renormalization of the model of incompressible fluid. 
Appendix~\ref{AppAdd} contains all calculations related to the renormalization
procedure and calculation of anomalous dimensions of the composite operators of advected field.

\section{Description of the model}
\label{sec:Model}

One of the possible approaches
to model fully developed turbulence within the framework of some microscopic model is to study 
the stochastic Navier-Stokes equation with a random external force~\cite{Legacy,LegacyD}. 
It has the form
\begin{equation}
\partial_{t} v_{k} + (v_{i} \partial_{i}) v_{k} + \partial_{k} \wp
= \nu_0\partial^2 v_{k} + \phi_{k},
\label{NS2}
\end{equation}
where 
$v_i(x)$ is a transverse (owing to the incompressibility) velocity field,
$x\equiv\left\{t,{\bm x}\right\}$,
$\partial _t \equiv \partial /\partial t$,
$\partial _i \equiv \partial /\partial x_{i}$,
$\nu _0$ is the molecular kinematic viscosity,
$\partial^2=\partial_i\partial_i$ is the Laplace operator,
$\wp= - \partial^{-2} (\partial_{i}v_{l})(\partial_{l} v_{i})$ 
is the pressure per unit mass, and
$\phi_{k}$ is the external force per unit mass.
Since the field $v_i(x)$ is incompressible we may use special units in which the density $\rho(x)=1$.
The turbulence is modeled by the force $\phi_{k}$, which is 
assumed to be a random variable. In stochastic formulation of the problem it mimics the input of
energy into the system
from the outer large scale $L$.
Without loss of generality the correlations of the random force $\phi_{k}$ 
in Fourier space read~\cite{Monin}
\begin{equation}
\label{ff}
\left\langle \phi_i(\omega,{\bm k})\ \phi_j(\omega',{\bm k}')\right\rangle
\propto \delta(\omega+\omega')\delta({\bm k}+{\bm k}')P_{ij}({\bm k})D_{\phi}(\omega,k),
\end{equation}
where $P_{ij}({\bm k})=\delta_{ij}-k_ik_j/k^2$ is the transverse projector 
and two $\delta$ functions are consequences of the translational invariance.

The Galilean invariance for the model requires the function $D_{\phi}(\omega,k)$ in Eq.~(\ref{ff}) to
be $\delta$ correlated in time~\cite{turbo}.
Nevertheless, 
it is very intriguing to consider such a model with a colored noise, i.e., with finite 
and not small correlation time, which is 
much more realistic from the physical point of view.
In general case this modification breaks the Galilean invariance, so the main
question of the paper is the following: is this symmetry restored 
in statistical sense for relevant measurable quantities?

The random force $\phi_{i}$ is simulated in the present paper by the statistical ensemble being a particular 
case of the Ornstein-Uhlenbeck process: it is assumed to be Gaussian and
homogeneous, with zero mean and correlation function~\cite{OU1,OU2,NonG}
\begin{eqnarray}
\label{VV-gen}
\left\langle \phi_i(t,{\bm x})\phi_j(t',{\bm x'})\right\rangle
= D_0\int\frac{d\omega}{2\pi}\int_{k>m}\frac{d{\bm k}}{(2\pi)^d}\ P_{ij}({\bm k}) 
\frac{k^{8-d-(y+2\eta)}}{\omega^2+\nu_0^2u_0^2k^{4-2\eta}}
\ e^{i{\bm k\cdot(\bm x-\bm x')}-i{\omega(t-t')}}.
\end{eqnarray}
Here $k \equiv |{\bm k}|$ is the wave number, 
$D_0>0$ is an amplitude factor,
$d$ is an arbitrary (for generality) dimension of ${\bm x}$ space, 
$1/m$ is the integral turbulence scale,
related to the stirring.
The function~(\ref{VV-gen}) involves
two independent exponents $y$ and $\eta$. The first one
describes the energy spectrum of the velocity in the inertial range 
${\cal E} \propto k^{1-y}$. The second exponent is related to the
dispersion law: 
the correlation time of the momentum $k$ scales as $k^{-2+\eta}$.
In the RG approach these two exponents play the role of two formal expansion parameters.
A new parameter $u_0$ is introduced for the dimensionality reasons.
Such ensemble was employed in some systems, studied
in Refs.~\cite{AKens,FinTime,FinTimeEta}. It was shown that depending on the values of
the exponents $y$ and $\eta$, the model reveals various types of
inertial-range scaling regimes with nontrivial anomalous exponents, which
were explicitly derived to the first~\cite{FinTime} and
second~\cite{FinTimeEta} orders of the double expansion in $y$ and $\eta$.

Depending on the parameter $u_0$, the function~(\ref{VV-gen}) demonstrates two
interesting
limiting cases: if $u_0\to0$, the situation corresponds to the
independent of time (``frozen'') random force, the case
$u_0\to\infty$ corresponds to the
zero-time correlated model.
The relations
\begin{equation}
\label{D0}
D_0/\nu_0^5 u_0^2=g_0\propto\Lambda^{y+2\eta}, \quad u_0\propto\Lambda^{\eta}
\end{equation}
define the coupling constant $g_0$, which plays the role of the
expansion parameter in the ordinary perturbation theory, and the
characteristic ultraviolet (UV) momentum scale $\Lambda$. 
The parameter $u_0$ introduced in Eq.~(\ref{VV-gen}) is written in the first expression for the
calculation reasons, in particular it is convenient in case of large $u_0$.

\section{Field theoretic formulation of the model}
\label{sec:QFT}

According to the general theorem~\cite{Zinn,Vasiliev,Tauber}, the stochastic problem~(\ref{NS2}) 
and~(\ref{VV-gen}) is equivalent to the field theoretic
model with a doubled set of fields $\Phi\equiv\left\{v_i', v_i\right\}$ and the De Dominicis-Janssen 
action functional, which can be written
in a compact form as
\begin{equation}
\label{Action}
{\cal S}_{ {\bm v}}(\Phi)= \frac{1}{2}v_i'D_{ik}v_k'+
v'_k\left[-\partial_t-(v_i\partial_i)
+\nu_0\partial^2\right]v_k.
\end{equation}
Here
$D_{ik}$ is the correlator~(\ref{VV-gen}) and
we employ a condensed notation, in which integrals over the spatial variable 
${\bm x}$ and the time variable $t$, as well as summation over the repeated indices, are omitted and
assumed implicitly
\begin{eqnarray}
\label{quadlocal2}
v'_k\partial_t{v}_k&=&\sum_{k=1}^d\int dt\int d{\bm x}\ v_k'(t,\bm x)\partial_t v_k(t,\bm x), \\ \nonumber 
v'_iD_{ik}v'_k&=&\sum_{i,k=1}^d\int dt \int dt' \int d{\bm x} \int {\mathrm d}{\bm x'}\ v_i(t,\bm x)D_{ik}(t-t',\bm x-\bm x')v_k(t',\bm x'). 
\end{eqnarray}
Since the auxiliary field $v'_k$ is transverse, i.e. $\partial_k v'_k=0$, the pressure term in expression~\eqref{Action} can be eliminated using integration by parts
\begin{eqnarray}
\label{Arpi}
v'_k\partial_k \wp=-\wp\partial_k v'_k=0.
\end{eqnarray}
Expressions~\eqref{Arpi} means that the field $v'_k$ acts as a transverse projector and selects the transverse parts of the expressions with which it is contracted.

\begin{figure}[t]
   \centering
   \begin{tabular}{c}
     \includegraphics*[width=7.5cm]{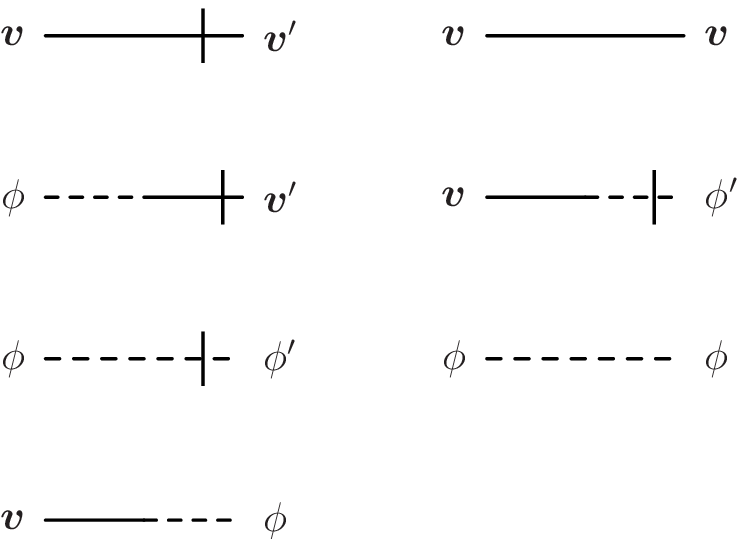}
      \end{tabular}
   \caption{Graphical representation of the bare propagators in the model~(\ref{Action}).}
   \label{fig:prop_vertex}
\end{figure}

\begin{figure}[b]
   \centering
   \begin{tabular}{c c}
     \includegraphics*[width=3.8cm]{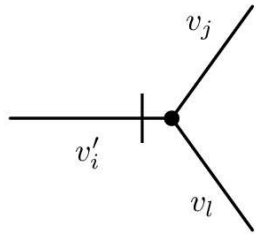}  
      \end{tabular}
   \caption{Graphical representation of the interaction vertex in the model~(\ref{Action}).}
   \label{fig:2}
\end{figure}

The field theoretic formulation~(\ref{Action}) means that various correlation 
and response functions of the original stochastic problem are
represented by functional averages over the full set of fields with the functional weight
$\exp {\cal S}_{ {\bm v}}(\Phi)$, and in this sense they can be
interpreted as the Green's
functions of the field theoretic model~\cite{Zinn,Vasiliev,Tauber}.
The perturbation theory of the model can be constructed according to the well-known 
Feynman diagrammatic expansion. 
Bare propagators are read off from the inverse matrix of the Gaussian (free) part of the action functional, while
a nonlinear part of the differential equation~\eqref{NS2} leads to the interaction vertex $-v_k'(v_i\partial_i)v_k$.
The propagator functions in 
the frequency-momentum representation read
\begin{eqnarray}
\label{VV}
\left\langle v_iv_{j}\right\rangle_0 &=&
D_0\
\frac{k^{8-d-(y+2\eta)}}{\omega^2+\nu_0^2u_0^2k^{4-2\eta}}\frac{P_{ij}({\bm k})}{\omega^2+\nu_0^2k^4},
\\
\label{VVprime}
\left\langle v_iv'_{j}\right\rangle_0 &=&
\frac{P_{ij}({\bm k})}{-i\omega+\nu_0k^2};
\end{eqnarray}
the triple vertex corresponds
to the expression
\begin{equation}
\label{Triple-vertex}
V_{ijl} = i\left(\delta_{il}k_j +\delta_{ij}k_l\right).
\end{equation}
Due to the incompressibility the derivative in the vertex can be moved onto the field $v_i'$, 
hence, $k_i$ is the momentum of the field $v'_i$.
A graphical representation of the propagator functions and interaction vertex are 
depicted in Fig.~\ref{fig:prop_vertex} and Fig.~\ref{fig:2}, respectively.
From now on, the end of a solid line without a slash denotes the field $v_i$, the end of a solid 
line with a slash denotes the field $v_i'$.

The analysis of UV divergences is based on the analysis of
the 1-irreducible Green's functions. 
In case of dynamical models~\cite{Vasiliev}
two independent scales (the time scale $T$ and the length scale $L$) have to be introduced: 
\begin{eqnarray}
[F] \sim [T]^{-d_{F}^{\omega}} [L]^{-d_{F}^{k}},
\label{canon}
\end{eqnarray}
where $d_{F}^{\omega}$ and $d_{F}^{k}$ are frequency and momentum dimensions of the quantity $F$, respectively.
The normalization conditions are 
\begin{eqnarray}
d_k^k=-d_{ x}^k=d_{\omega }^{\omega }=-d_t^{\omega }&=1, \\ \nonumber
d_{\omega }^k=d_t^k=d_k^{\omega} =d_{x}^{\omega }&=0.
\end{eqnarray}
Based on $d_F^k$ and $d_F^\omega$ the total canonical dimension
$d_F=d_F^k+2d_F^\omega$ can be introduced, which in
the renormalization theory of the dynamic models plays the
same role as the conventional (momentum) dimension does in the
static problems. 

The canonical dimensions for the model~(\ref{Action}) are given in
Table~\ref{table1}, including the renormalized parameters (without the subscript ``$0$''), which will be
introduced later. 
The parameters $\theta$, $\theta'$, $\kappa$, and $w$ are connected with the problem of the advection 
of scalar field and will be used in Secs.~\ref{sec:adv}~-- \ref{sec:final}. From Table~\ref{table1} it
follows that our model
is logarithmic (the coupling constants $g_{0} \sim [L]^{-y}$
and $u_{0} \sim [L]^{-\eta}$ are simultaneously dimensionless) at $y=\eta=0$, so in the minimal
subtraction (MS) scheme, which we use in this paper,
the UV divergences in the Green's functions manifest
themselves as poles in $y$, $\eta$ and their linear combinations.

The total canonical dimension of an arbitrary 1-irreducible Green's function
$\Gamma_{N_\Phi} = \langle\Phi \dots \Phi \rangle _{\rm 1-ir}$ is given by
the relation
\begin{equation}
d_{\Gamma}= d+2- \sum_{\Phi} N_{\Phi }d_{\Phi} = d+2-
 N_{v'} d_{v'} - N_{v} d_{v}.
\label{dGamma}
\end{equation}
Here $N_{\Phi}=\{N_{v},\,N_{v'}\}$ are the numbers of
corresponding fields entering the function $\Gamma$ and
$d_{\Phi}$ is the corresponding total canonical dimension of a field 
$\Phi$~\cite{Zinn,Vasiliev,Tauber}.
The superficial UV divergences the removal of which requires counterterms might be
present only in
the functions $\Gamma$ for which
the formal index of divergence $\delta_{\Gamma}$ (being the value of $d_{\Gamma}$ in the logarithmic theory) is
a non-negative integer.
Dimensional analysis should be augmented by the following considerations:

(1) In any dynamical model of the type~(\ref{Action}) all the 1-irreducible functions without the
response field $v_i'$
necessarily contain closed circuits of retarded propagators similar to~\eqref{VVprime}. Therefore, such functions
vanish identically and do not require counterterms.

(2) Using the transversality condition of the field $v_i$ we can move the derivative in the
vertex $-v_k'(v_i\partial_i)v_k$ from the field $v_k$ onto the field $v_i'$.
Therefore, in any 1-irreducible diagram it is always possible to move
derivatives onto external ``tails'' $v_i'$, which reduces the real
index of divergence: $d'_{\Gamma} = d_{\Gamma} - N_{v'}$. 

\begin{table}[t]
\caption{Canonical dimensions of the fields and parameters in the model~(\ref{Action}).}
\label{table1}
\begin{ruledtabular}
\begin{tabular}{c||c|c|c|c|c|c|c|c|c}
$F$ & $\boldsymbol{v}' $ & $\boldsymbol{v}$ & $\theta$ & $\theta'$ & $M,m,\mu, \Lambda $ &
$\nu ,\nu_{0}$, $\kappa$, $\kappa_0$ & $u, u_0$ 
& $g_{0}$ & $g$, $w$, $w_0$ \\
\hline
$d_{F}^{\omega}$     & $-1$    & $1$    & $-1/2$       & 1/2   & $0$   & $1$              & 0      & 0   & 0 \\
\hline
$d_{F}^{k}$           & $d+1$   & $-1$   & 0  & $d$   & $1$   & $-2$       & $\eta$ & $y$ & 0 \\
\hline
$d_{F}$               & $d-1$   & $1$    & $-1$    & $d+1$ & $1$   & $0$          & $\eta$ & $y$ & 0 \\
\end{tabular}
\end{ruledtabular}
\end{table}

From Table~\ref{table1} and Eq.~(\ref{dGamma}) we find that
\begin{equation}
d_{\Gamma}=  d+2- d N_{v'} - N_{v}.
\label{dGammaPrime}
\end{equation}
From this expression we conclude that for any $d>2$, superficial
divergences can be present only in the 1-irreducible functions of two types. The first example is the function
$\langle v_\alpha'v_\beta\rangle_{\rm 1-ir}$, for which the real index of divergence is $d_{\Gamma}=2$. Another
possibility is the function
$\langle v_\alpha'v_\beta v_\gamma\rangle_{\rm 1-ir}$ with $d_{\Gamma}=1$. This means, that all
the UV divergences in our model can be removed by the counterterms of the form $v_i\partial^2v_i$ 
and $v'_k(v_i\partial_i)v_k$.
The Galilean invariance which holds in case of $\delta$ correlated in time function
$D_\phi(\omega,k)$ forbids the divergence of the vertex 
$\langle v_\alpha'v_\beta v_\gamma\rangle_{\rm 1-ir}$, which we have in our case of the colored
noise~(\ref{VV-gen}).
For $d = 2$ a new UV divergence arises in the function $\langle v'_\alpha v'_\beta\rangle_{\rm 1-ir}$, and a new
counterterm $v_i'\partial^2v_i'$ should be included~\cite{Two}. This case requires a 
special treatment, and in the following we assume $d>2$.

The model~(\ref{Action}) is multiplicatively renormalizable with two independent
renormalization constants $Z_{1}$ and $Z_{2}$; the renormalized action functional has the form
\begin{equation}
\label{Action-Ren}
{\cal S}_{{ {\bm v}}_R}(\Phi)= \frac{1}{2}v_i'D_vv_k'+
v'_k\left[-\partial_t-Z_1(v_i\partial_i)
+Z_2\nu\partial^2\right]v_k.
\end{equation}
Here $g$, $\nu$, and $u$ are the renormalized counterparts of the original (bare)
parameters, the function $D_{v}$ is expressed
in the renormalized parameters using the relation $g_{0}\nu_0^{5}u_0^{2} =
g\mu^{y+2\eta}\nu^5u^2$; the reference scale $\mu$
is an additional free parameter of the renormalized theory.

The renormalized action~(\ref{Action-Ren}) is obtained from the original one~(\ref{Action})
by the renormalization of the fields $v\to Z_{v}v$, $v'\to Z_{v'}v'$ and the parameters
\begin{equation}
g_{0}=g\mu^{y}Z_{g}, \quad u_{0}=u\mu^{\eta}Z_{u}, \quad \nu_{0}=\nu Z_{\nu}.
\label{mult}
\end{equation}
The renormalization constants in Eqs.~(\ref{Action-Ren}) and~(\ref{mult}) are
related as
\begin{eqnarray}
Z_{\nu} = Z_{2}, \quad Z_{u}=Z_{2}^{-1}, \quad
Z_{g} = Z_{1}^2Z_2^{-3}, \quad Z_{v}=Z_{v'}^{-1}=Z_{1}.
\label{relat}
\end{eqnarray}
The renormalization constants are found from the requirement that the
Green's functions of the renormalized model~(\ref{Action-Ren}), when expressed
in renormalized variables, have to be UV finite and can
depend only on the completely
dimensionless parameters $g,u,d,y$, and $\eta$. 

\section{Renormalization of the model and RG functions}
\label{RenModel}

Let us consider 
the generating functional of the 1-irreducible Green's
functions 
\begin{equation}
\Gamma(\Phi) = {\cal S}_{ {\bm v}}(\Phi) + \widetilde{\Gamma}(\Phi),
\end{equation}
where ${\cal S}_{ {\bm v}}(\Phi)$ is the action functional~(\ref{Action}) and
$\widetilde{\Gamma}(\Phi)$ is the sum of all the 1-irreducible diagrams
with loops.

\begin{figure}[t]
\center
\includegraphics[width=.3\textwidth,clip]{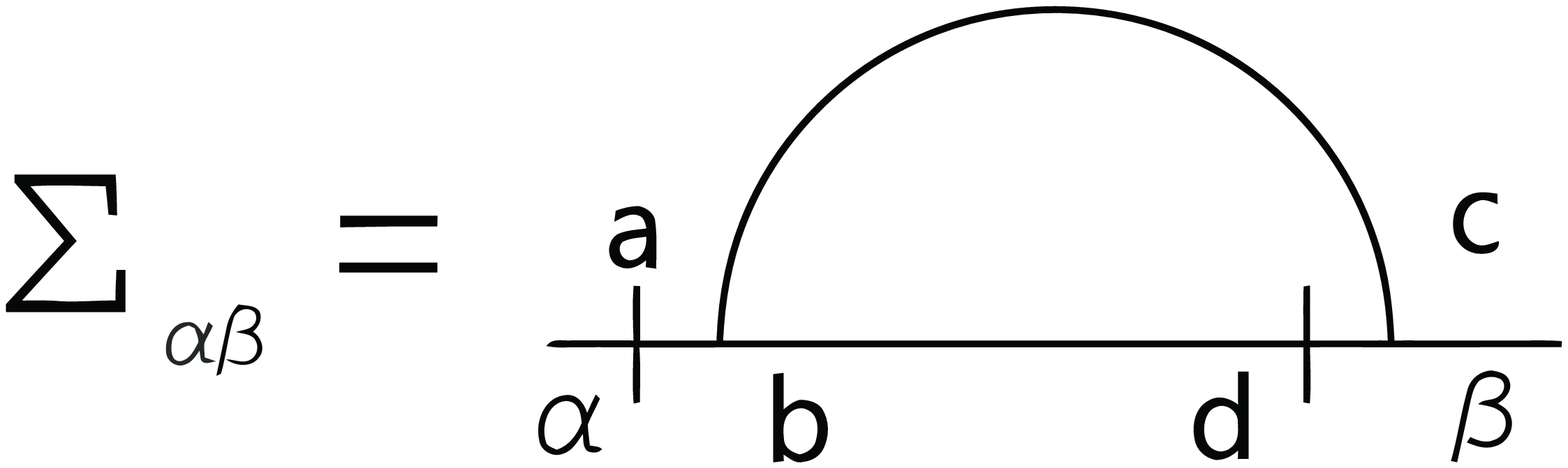}
\caption{The one-loop approximation of the 1-irreducible response function 
$\langle v_{\alpha}^{\prime}v_{\beta}\rangle_{\rm 1-ir}$. 
Greek letters denote external (free) indices of the diagram, Latin
letters correspond to internal indices of the vector fields with implied summation.}
\label{fig:SelfEnergy}
\end{figure}

\begin{figure}[b]
\center
\includegraphics[width=.5\textwidth,clip]{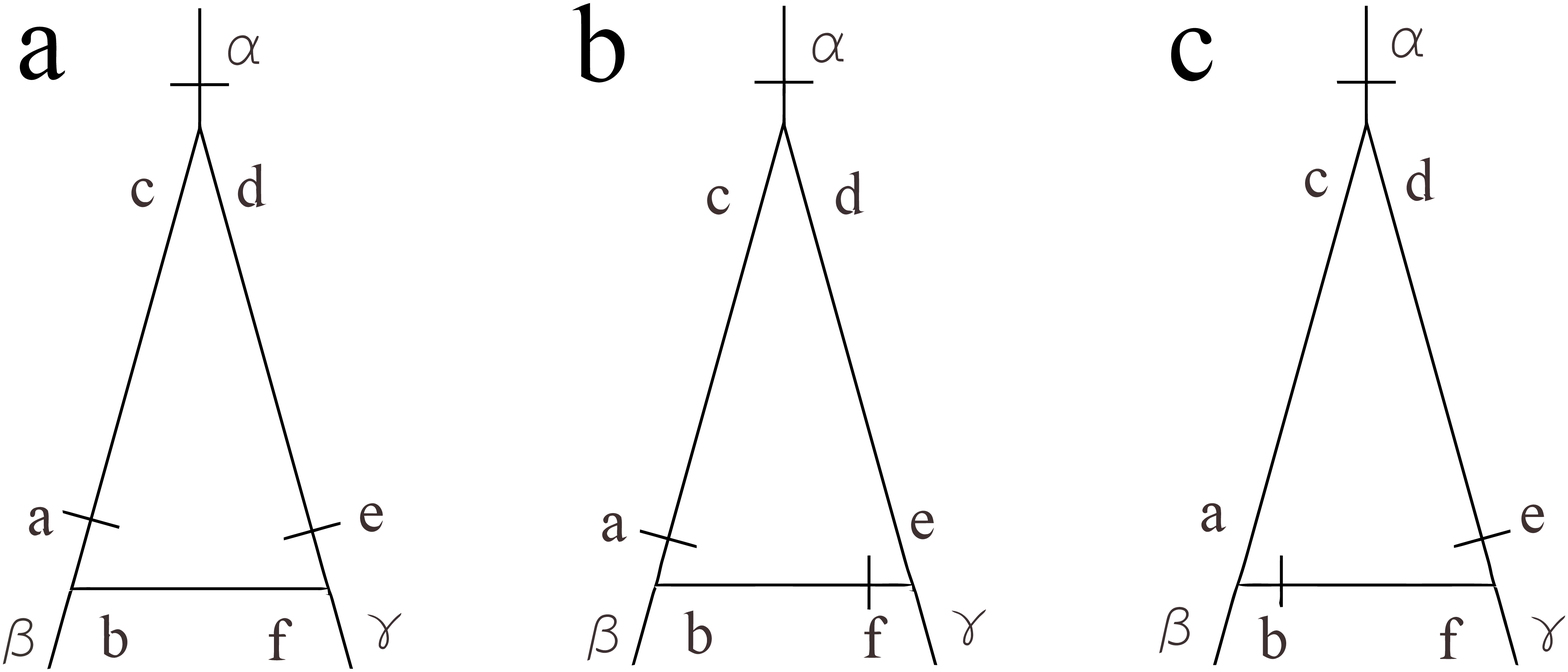}
\caption{The one-loop approximation of the 1-irreducible function 
$\langle v_\alpha'v_\beta v_\gamma\rangle_{\rm 1-ir}$. 
Greek letters denote external (free) indices of the diagram, 
Latin letters correspond to internal indices of the vector fields with implied summation.}
\label{fig:Trio}
\end{figure}

Hence, one-loop approximation for the 1-irreducible Green's functions that require UV renormalization provides
\begin{eqnarray} 
\label{Dyson}
\langle v_\alpha'v_\beta \rangle_{\rm 1-ir}&=&\left(i\omega -\nu_0 p^2\right) P_{\alpha\beta}({\bm p})+\Sigma_{\alpha\beta}, \\
\label{VVV}
\langle v_\alpha'v_\beta v_\gamma\rangle_{\rm 1-ir}&=&V_{\alpha\beta\gamma}+(\Delta_1+\Delta_2+\Delta_3).
\end{eqnarray}
Here, $P_{\alpha\beta}({\bm p})$
is the transverse projector, $\Sigma_{\alpha\beta}$ is the
self-energy operator, graphical representation for which is depicted in the
Fig.~\ref{fig:SelfEnergy}, $V_{\alpha\beta\gamma}$ is the vertex factor~(\ref{Triple-vertex}),
 and diagrams $\Delta_1$, $\Delta_2$, and $\Delta_3$ are depicted in 
Figs.~\ref{fig:Trio}a--\ref{fig:Trio}c.

The calculation of the renormalization constants $Z_1$ and $Z_2$ in
the one-loop approximation gives 
\begin{eqnarray} 
\label{Z1}
Z_1&=&1-g\frac{1}{d(d+2)}\frac{u}{(u+1)^3}\frac{1}{y},\\
Z_2&=&1-g\frac{1}{d(d+2)}\frac{u^3d(d-1)+3u^2d(d-1)+2u(d^2-d+2)}{4(u+1)^3}\frac{1}{y}.
\label{Z2}
\end{eqnarray}
Here, we have introduced a new coupling constant $g\to g S_d/(2\pi)^d$
with $S_d$ being the surface area of the unit sphere in $d$-dimensional space, 
see Appendixes~\ref{a1} and~\ref{a2} for details.
The corrections of orders $g^2$ and higher are neglected.

The relation between the initial and renormalized action functionals $S(\Phi,e_{0})= S^{R}(Z_\Phi\Phi,e,\mu)$, 
where $e$ is the complete set of parameters, yields the fundamental RG differential equation:
\begin{equation}
\biggl\{ \widetilde{\cal D}_{\mu} + N_{v}\gamma_{v} +
N_{v'}\gamma_{v'} \biggr\} \,G^{R}(e,\mu,\dots) = 0,
\label{RG1}
\end{equation}
where $G =\langle \Phi\cdots\Phi\rangle$ is the correlation function of the fields~$\Phi$;
$N_{v}$ and $N_{v'}$ are the numbers of the renormalization-requiring fields ${\bm v}$ and ${\bm v'}$,
 respectively, which are the inputs to
$G$; the ellipsis in the expression~(\ref{RG1}) stands for the other arguments of $G$ (spatial
and time variables, etc.).
Further, $\widetilde{\cal D}_{\mu}$ is the differential operation $\mu\partial_{\mu}$ taken for fixed
$e_{0}$ and expressed in the renormalized variables:
\begin{equation}
\widetilde{\cal D}_{\mu}= {\cal D}_{\mu} + \beta_{g}\partial_{g} + \beta_{u}\partial_{u} 
- \gamma_{\nu}{\cal D}_{\nu}.
\label{RG2}
\end{equation}
Here and below we have denoted ${\cal D}_{x} \equiv x\partial_{x}$ for any variable $x$.
The anomalous dimension $\gamma_{F}$ of a certain quantity $F$
(a field or a parameter) is defined as
\begin{equation}
\gamma_{F}= Z_{F}^{-1} \widetilde{\cal D}_{\mu} Z_{F} =
\widetilde{\cal D}_{\mu} \ln Z_F.
\label{RGF1}
\end{equation}
The $\beta$ functions for the two dimensionless coupling
constants $g$ and $u$ are
\begin{eqnarray}
\beta_{g} &=& \widetilde{\cal D}_{\mu} g = g(-y-\gamma_{g}),
\nonumber \\
\beta_{u} &=& \widetilde{\cal D}_{\mu} u = u(-\eta-\gamma_{u}),
\label{betagw}
\end{eqnarray}
where the latter equations result from the definitions and the
relations~(\ref{mult}).

From the definitions and expressions~(\ref{Z1})~-- (\ref{Z2}) for the renormalization 
constants $Z_1$ and $Z_2$ one finds
\begin{eqnarray}
\label{gamma}
\gamma_1 &=&
g\frac{1}{d(d+2)}\frac{u}{(u+1)^3},\\
\gamma_2 &=&
g\frac{1}{d(d+2)}\frac{u^3d(d-1)+3u^2d(d-1)+2u(d^2-d+2)}{4(u+1)^3},
\end{eqnarray}
and from the relations~(\ref{relat}) it follows that
\begin{equation}
\label{Beta-zero}
\beta_g=g(-y-2\gamma_1+3\gamma_2),\quad\beta_u=u(-\eta+\gamma_2).
\end{equation}
Eqs.~\eqref{Beta-zero} give us full set of $\beta$ functions defining fixed points, which are responsible for asymptotic behavior of correlation and structure functions.

\section{IR attractive fixed points}
\label{sec:RC}

One of the basic RG statements is that the asymptotic behavior of the model
is governed by the fixed points $\left\{g^*,u^*\right\}$, 
defined by the equations $\beta_g=0$, $\beta_u=0$. 
The type of a fixed point (IR/UV attractive or a saddle point), i.e., the
character of the RG flow in the vicinity of the point, is determined by the
matrix $\Omega_{ik} = \partial\beta_i/\partial g_k$ at a given point, where $\beta_i$ is
the full set of $\beta$ functions and $g_k$ is the full set of couplings.
For an IR attractive fixed point, the matrix $\Omega$ has to be positive definite,
i.e., the real parts of all its eigenvalues have to be positive.

A direct analysis of the $\beta$ functions reveals the
existence of the three fixed points: the trivial one and two nontrivial.
The free Gaussian fixed point, for which all interactions are irrelevant and
no scaling and universality are expected, has the coordinates
\begin{equation}
\label{Triv}
g^*=0,\quad u^*=0
\end{equation}
and is IR attractive if both $y$ and $\eta$ are negative.
 
\begin{figure}[t]
\center
\includegraphics[width=.40\textwidth,clip]{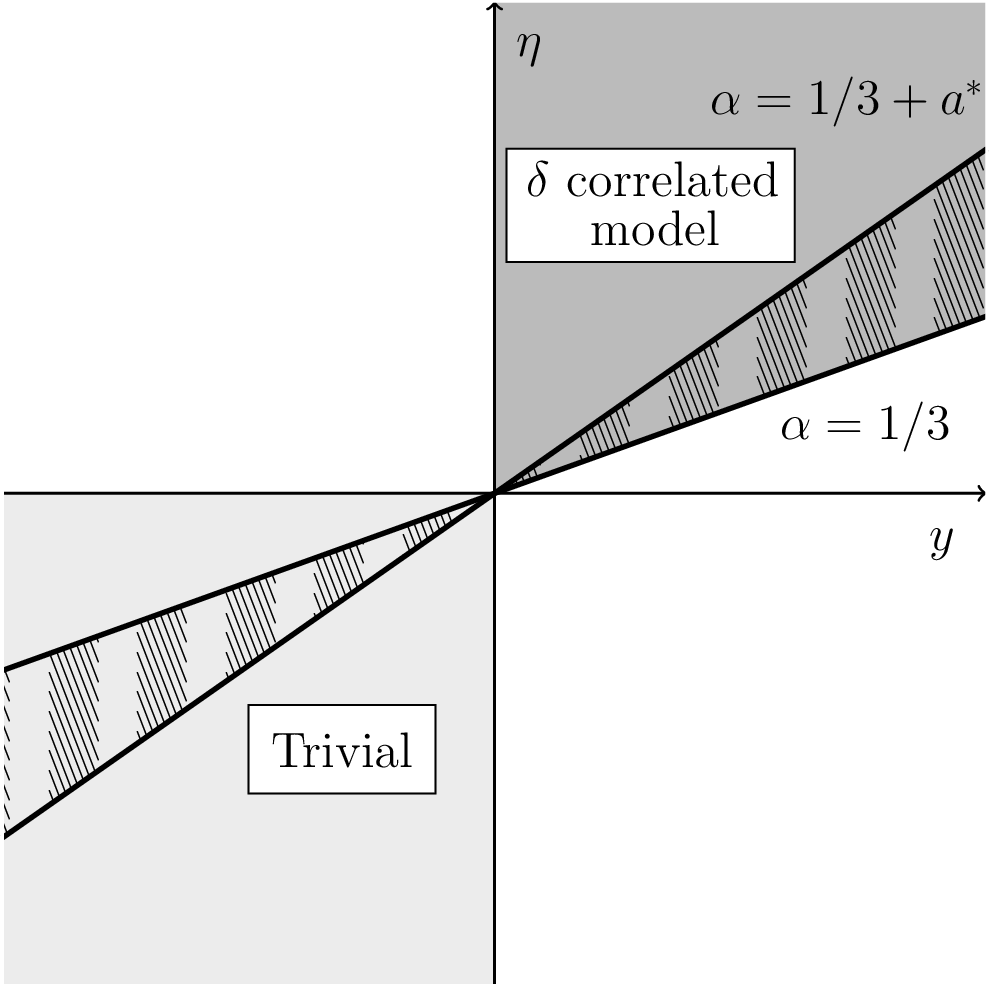}
\caption{Domains of the existence and IR stability of the fixed points for the model
(\protect\ref{Action}) in the plane $(y,\eta)$; $a^*=\frac{4/3}{3d(d-1)+2}$.
Gray areas correspond to the regions where the trivial and $\delta$ correlated model fixed points
are IR attractive;
hatched areas which are lying inside the gray ones correspond to the regions where another 
nontrivial (saddle type) point exists;
in white areas there is no IR attractive fixed point.}
\label{fig:FP}
\end{figure}

Let us we define $\alpha\equiv\eta/y$ [see Eq.~(\ref{VV-gen})]. 
If parameter $\alpha$ satisfies the inequalities
\begin{equation}
\label{ineq}
\frac{1}{3}<\alpha<\frac{1}{3}+\frac{4/3}{3d(d-1)+2},
\end{equation}
the system possesses the fixed point $\left\{g^*,u^*\right\}$ with the coordinates 
\begin{eqnarray}
\label{Saddle-1}
u^* &=&
\frac{-3+\sqrt{1-\frac{16(\alpha-1)}{d(d-1)(3\alpha-1)}}}{2},
\\
\label{Saddle-2}
g^* &=&
d(d+2)\frac{(u^*+1)^3}{u^*}\frac{3\alpha-1}{2}y,
\end{eqnarray}
see Fig.~\ref{fig:FP}.
However, it turns out that one of the two eigenvalues of the matrix $\Omega$ for this point is
negative. This means, that this fixed point is a saddle one, i.e., 
for any values of $y$ and $\eta$ it is IR attractive only in one of the two possible directions. 
This fixed point exists for all $d$ except the limit $d\to\infty$, where the inequality~(\ref{ineq}) has
no solution (see details below).

Another case to be considered is $u^*\to\infty$. 
From Eqs.~(\ref{VV-gen}) and~(\ref{VV}) it follows that this case corresponds to the 
previously studied model with the $\delta$ correlated in time random force~\cite{NS-Zero}. 
Therefore, one should obtain the well-known fixed point of this model.
This is indeed the case. To prove this statement
it is convenient to pass from the
variable $u$ to a variable $x = 1/u$.
The limit $u\to\infty$ corresponds now to the limit $x\to 0$; the new $\beta$
function is
\begin{eqnarray}
\label{bux}
\beta_{x} &=& \widetilde{\cal D}_{\mu} x = -\frac{1}{u^2}\beta_{u}.
\end{eqnarray}
If $u^*\to\infty$ anomalous dimensions have the following simple form
\begin{eqnarray}
\gamma_1 &=& 0, \\
\gamma_2 &=& g\frac{d-1}{4(d+2)}.
\end{eqnarray}
Therefore, we obtain the new set of $\beta$ functions
\begin{eqnarray}
\label{betaInf}
\beta_{g} &=& g\left[-y+g\frac{3(d-1)}{4(d+2)}\right], \\
\beta_{x} &=& x\left[\eta-g\frac{d-1}{4(d+2)}\right].
\label{betaInf2}
\end{eqnarray}
From Eqs.~(\ref{betaInf})~-- (\ref{betaInf2}) it follows that the system possesses the fixed point
with the coordinates 
\begin{equation}
\label{IRattr}
x^*=0, \quad g^* = \frac{4(d+2)}{3(d-1)}y,
\end{equation}
which coincides with the results of Ref.~\cite{NS-Zero} and 
is IR attractive for $y>0$ and $\eta>y/3$.

An interesting situation corresponds to the limit $d\to\infty$. 
The study of the large $d$ behavior of the fluid turbulence is not only an academic interest: 
one can hope that in this case the intermittency and anomalous scaling disappear or
acquire a simple ``calculable'' form and the finite-dimensional turbulence can be studied within
the expansion around this ``solvable'' limit; hence the idea of expansion in $1/d$, considered 
in Refs.~\cite{InfD1,InfD2,InfD3,InfD4}.
If $d\to\infty$ the set of the $\beta$ functions~(\ref{Beta-zero}) reads
\begin{eqnarray}
\label{betaDInf}
\beta_{g} &=& g\left[-y+g\frac{3u(u+2)}{4(u+1)^2}\right], \\
\beta_{u} &=& u\left[-\eta+g\frac{u(u+2)}{4(u+1)^2}\right].
\label{betaDInf2}
\end{eqnarray}
Therefore, the system $\beta_g=0$, $\beta_u=0$ 
admits several possible solutions. The trivial one is $g^*=u^*=0$; it corresponds to Eqs.~(\ref{Triv}) 
at finite $d$.
Another solution is an infinite fixed point
\begin{equation}
\label{IRattr4}
x^*=0, \quad g^* = \frac{4}{3}y,
\end{equation}
where $x=1/u$ (see above) and $\beta_x$ is given by expression~\eqref{bux}.
This point is IR attractive for $y>0$ and $\eta>y/3$ and
corresponds to Eqs.~(\ref{IRattr}). 
Furthermore, there is one more solution of the system~\eqref{betaDInf}~-- \eqref{betaDInf2}:
\begin{equation}
\label{IRattr2}
g^*\frac{u^*(u^*+2)}{4(u^*+1)^2} = \frac{y}{3}=\eta,
\end{equation}
where both $g^*$ and $u^*$ are undefined separately. This case corresponds to the saddle type
point~(\ref{Saddle-1})~-- (\ref{Saddle-2}), but with one significant difference: if $d\to\infty$, two 
eigenvalues of the matrix $\Omega$ are
\begin{equation}
\label{IRattr3}
\lambda_1 =0; \quad \lambda_2=\left.g\frac{u(3u^2+9u+8)}{4(u+1)^3}\right|_{g^*,u^*}.
\end{equation}
Eqs.~(\ref{IRattr2}) and~(\ref{IRattr3}) are in agreement with the results~(\ref{Saddle-1})~-- (\ref{Saddle-2}) 
for finite $d$. If $d\to\infty$, there is no solution for inequality~(\ref{ineq}), and two hatched 
triangles, denoting the possible areas of existence of this point, degenerate into one line
$\eta=\frac{y}{3}$ (see Fig.~\ref{fig:FP}). At the same time one of the two eigenvalues, which was
negative at finite $d$, tends to zero. 
This means, that we have not a point, but a line 
$\eta=\frac{y}{3}$ with zero velocity along it, 
which is (and was at finite $d$) IR attractive in the perpendicular direction. 
It is very intriguing phenomenon. On the one hand, there is a continuous limit from finite $d$ to the
case $d\to\infty$. 
Moreover, the expression $\lim_{d\to\infty}\lambda_1=0$ can be checked directly from the original
$\beta$ functions at the fixed point~(\ref{Saddle-1})~-- (\ref{Saddle-2}). On the other hand, the 
saddle type point, which can be reached only if the initial data (both position and velocity) are
very specific and allow it, transforms into an IR attractive line, which will be achieved for any
initial data.

The results presented in this section are based on the explicit form
of the $\beta$ functions \eqref{gamma}~-- \eqref{Beta-zero} derived within the leading one-loop
approximation. They exhibit a very fine structure and can well appear
to be sensitive to inclusion of higher-order corrections.
This interesting issue lies beyond the scope of our discussion and
will be a subject of further investigation.

\section{Critical dimensions}
\label{Sec:CD}

In the leading order of the IR asymptotic behavior, the Green's functions 
satisfy the RG equation with the substitution
$g\to g^{*}$ and $u\to u^{*}$. This property together with canonical
scale invariance gives us the critical dimensions of the fields in the model,
which, in fact, govern the asymptotic behavior of arbitrary correlation 
functions
\begin{equation}
\Delta_{F} = d^{k}_{F}+ \Delta_{\omega}d^{\omega}_{F} + \gamma_{F}^{*},
\quad {\rm where}
\quad \Delta_{\omega} = - \Delta_{t} = 2-\gamma_{\nu}^{*}.
\label{Krit}
\end{equation}
Here, $\Delta_{F}$ denotes the critical dimension of 
the quantity $F$, while $\Delta_{t}$ and $\Delta_{\omega}$ are
the critical
dimensions of time and frequency. The symbol $\gamma_{F}^{*}$ denotes the value $\gamma_{F}$ at the fixed point.

If $u^*\to\infty$ one obtains the exact answers (with no corrections of orders $y^2$ and higher)
\begin{equation}
\Delta_v = 1-y/3, \quad \Delta_{v'} = d- 1+y/3,
\end{equation}
which are in agreement with Ref.~\cite{NS-Zero}.

The saddle fixed point~(\ref{Saddle-1})~-- (\ref{Saddle-2}) gives 
\begin{equation}
\Delta_v = 1+\frac{\eta-y}{2}, \quad \Delta_{v'} = d-1+\frac{\eta-y}{2}.
\end{equation}
The latter dimensions have higher-order corrections $y^n \eta^m $ with $n+m \geqslant 2$. Evaluation 
of these corrections requires the consideration beyond the one-loop approximation.

\section{Advection of passive scalar fields}
\label{sec:adv}

Let us consider a passive advection of a scalar field $\theta(x)\equiv \theta(t,{\bm x})$, which 
satisfies the stochastic differential equation
\begin{equation}
\partial _t\theta+ \partial_{i}(v_{i}\theta)=\kappa _0 \partial^{2} \theta + f_\theta.
\label{density1}
\end{equation}
Eq.~(\ref{density1}) describes the advection of a density field, e.g., density of a pollutant.
The advection of a tracer field (temperature, specific entropy, or concentration of the impurity 
particles) differs from this case by the transformation $\partial_{i}(v_{i}\theta)\to(v_{i}\partial_{i})\theta$ 
on the left-hand side of Eq.~(\ref{density1}). Therefore, in case of an incompressible carrier flow
(i.e., if $\partial_i v_i=0$) both density and tracer fields are described by the same equation.

The newly introduced coefficient
$\kappa_0$ is the molecular
diffusivity, $\partial^{2}=\partial _i\partial _i$
is the Laplace operator, $v_i(x)$ is the velocity field, which obeys Eq.~(\ref{NS2}),
and $f_\theta= f_\theta(x)$ is a Gaussian noise with zero mean and given covariance
\begin{equation}
\langle f_\theta (x)f_\theta (x') \rangle = \delta(t-t')\, C({\bm r}/L_\theta), \quad
{\bm r}= {\bm x} - {\bm x}'.
\label{noise}
\end{equation}
The function $C({\bm r}/ L_\theta)$ in Eq.~(\ref{noise}) is finite at $({\bm r}/L_\theta)\to 0$ and rapidly
vanishes when $({\bm r}/L_\theta)\to\infty$. The expression~(\ref{noise}) brings about another
external (integral) scale $L_\theta$,
related to the noise $f_\theta$, but henceforth we will not distinguish it from
its analog $L=m^{-1}$ in the correlation function of the stirring force~(\ref{VV-gen}). 
The noise simulates effects of initial and boundary conditions of the system.

If the velocity $v_i$ obeys the stochastic Navier-Stokes equation~(\ref{NS2}), the
problem~(\ref{density1}),~(\ref{noise}) is tantamount
to the field theoretic model of the full set of fields
$\widetilde{\Phi}\equiv\{\theta', \theta, v_i', v_i\}$
and the action functional
\begin{equation}
{\cal S}(\widetilde{\Phi})= {\cal S}_{\theta}(\theta', \theta, v_i) + {\cal S}_{ {\bm v}}(v_i', v_i),
\label{Fact}
\end{equation}
where the advection-diffusion component
\begin{eqnarray}
{\cal S}_{\theta} (\theta', \theta, v_i) = \frac{1}{2} \theta' D_{f} \theta' +
\theta' \left[ - \partial_{t}\theta -
\partial_{i}(v_{i}\theta) +\kappa _0 \partial^{2} \theta \right]
\label{Dact}
\end{eqnarray}
is the De Dominicis-Janssen action for the stochastic problem~(\ref{density1}), (\ref{noise})
at fixed $v_i$, while the second term is given 
by~(\ref{Action}) and represents the velocity statistics; $D_{f}$ is the
correlation function~(\ref{noise}), all the required
integrations and summations over the vector indices are assumed, see explanations~(\ref{quadlocal2}).

In addition to the expressions~(\ref{VV})~-- (\ref{Triple-vertex}), the diagrammatic technique in the full
problem involves a new vertex $-\theta'\partial_{j}(v_{j}\theta)$ and two new propagators:
\begin{eqnarray}
\left\langle \theta \theta' \right\rangle _0 &=&
\frac{1} {-i\omega +\kappa _0 k^2}, \\
\left\langle \theta \theta \right\rangle _0 &=& \frac {C({\bm k})}
{\omega^{2} +\kappa_0^{2} k^4}.
\label{lines3}
\end{eqnarray}
\begin{figure}[b]
   \centering
    \begin{tabular}{c}
     \includegraphics[width=6.9cm]{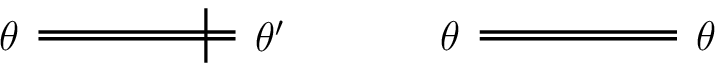}
    \end{tabular}
   \caption{Graphical representation of the bare propagators $\langle \theta \theta' \rangle _0$ and $\langle \theta \theta \rangle _0$.}
  \label{P2}
\end{figure}

In the frequency-momentum representation the new vertex reads 
\begin{eqnarray}
V_{j} ({\bm k}) = ik_{j},
\label{vertex1}
\end{eqnarray}
where ${\bm k}$ is the momentum carried by the field $\theta'$.

\begin{figure}[t]
   \centering
    \begin{tabular}{c}
     \includegraphics*[width=3.5cm]{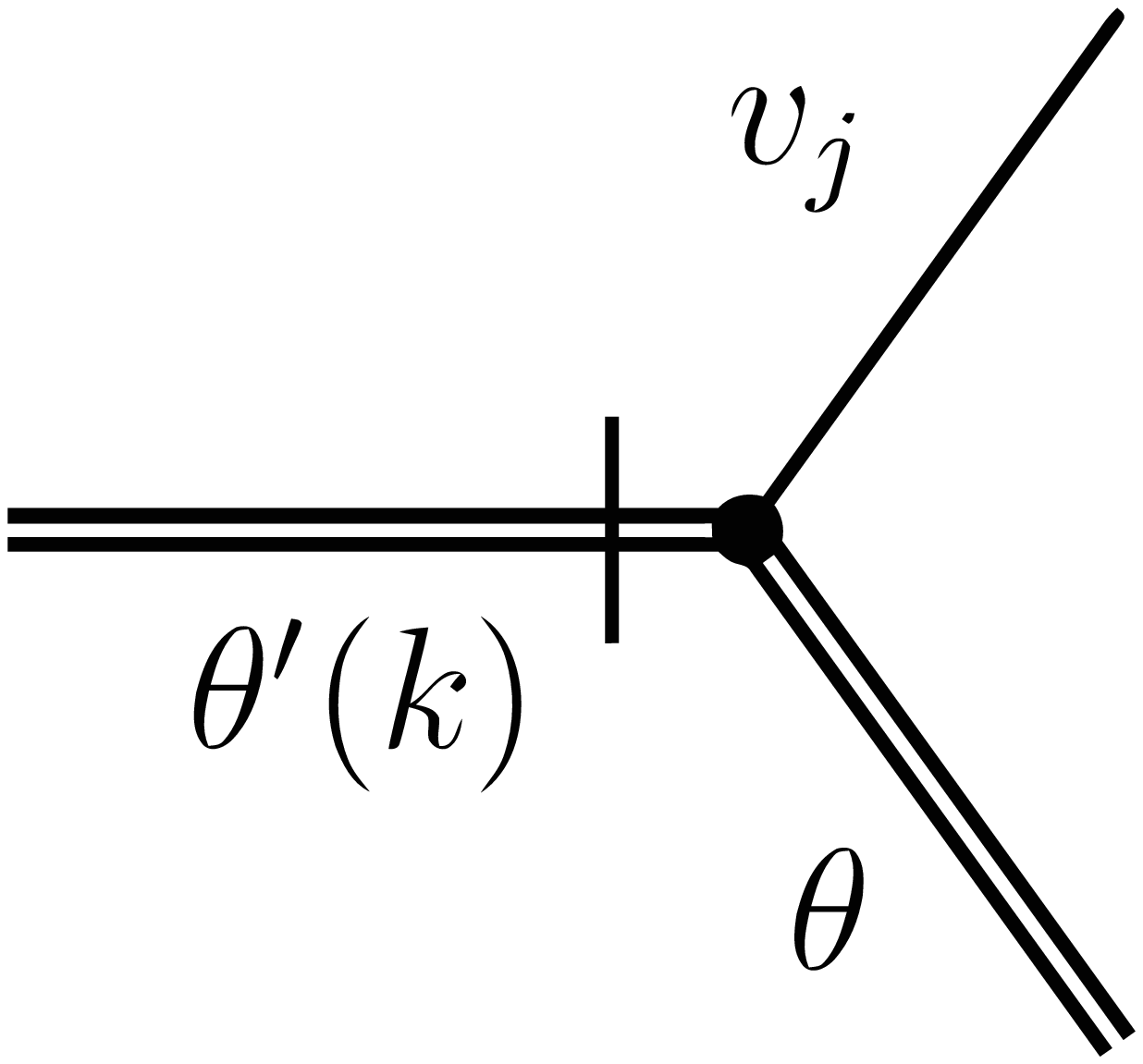}
    \end{tabular}
  \caption{Graphical representation of the interaction vertex $V_j$.}
 \label{V3}
\end{figure}

Graphical representations of the newly introduced propagator functions and vertex are depicted
in Figs.~\ref{P2} and~\ref{V3}, respectively.
From now on, the end of a double solid line without a slash denotes the field $ \theta $, the end
of a double solid line with a slash denotes the field $ \theta'$.

The model~(\ref{Fact}) was considered earlier in Ref.~\cite{AVH84} in case of the zero-time correlation 
function 
\begin{eqnarray}
\left\langle \phi_i(t,{\bm x})\phi_j(t',{\bm x'})\right\rangle
= \delta(t-t')\widetilde{D}_0\int_{k>m}\frac{d{\bm k}}{(2\pi)^d}\ P_{ij}({\bm k})k^{4-d-y}\ e^{i{\bm k\cdot(\bm x-\bm x')}},
\end{eqnarray}
where $\widetilde{D}_0=D_0/\nu_0^2u_0^2$.
Here, we consider the Navier-Stokes equation~(\ref{NS2}) with the colored
random force~(\ref{VV-gen}). 
As it was shown above the only IR attractive fixed point~(\ref{IRattr}) of this system corresponds to
the model with zero-time correlations. This means, that the Galilean symmetry broken by colored noise is restored and we take advantage
of the previous study, namely
the fact that the full model~(\ref{Fact}) is multiplicatively renormalizable and possesses the IR 
attractive fixed point $\left\{g^*, w^*\right\}$ in the full space of couplings
\begin{eqnarray}
\label{FP}
g^* = \frac{4(d+2)}{3(d-1)}y+O\left(y^2\right), \quad w^*(w^*+1) = \frac{2(d+2)}{d}.
\end{eqnarray}
Here, we have introduced the new dimensionless variable $w_0=\kappa_0/\nu_0$ with $\nu_0$ from Eq.~(\ref{NS2})
and its renormalized analog $w$. The critical dimensions of the advected field $\theta$ and 
additional field $\theta'$ are
\begin{equation}
\label{ThDim}
\Delta_\theta = -1+y/6, \quad \Delta_{\theta'} = d+ 1-y/6
\end{equation}
and have no corrections of the orders $y^2$ and higher.

\section{Operator product expansion for the pair correlation function and large-scale anisotropy}
\label{sec:ope}

Let us consider the influence of the large-scale anisotropy, introduced into the system at the 
large scale $L$ through the correlation function of the random noise~(\ref{noise}), on the inertial
range behavior of the pair correlation function
$\langle\theta(t,{\bm x}) \theta(t,{\bm x}') \rangle$. The goal is to check Kolmogorov's local
symmetry restoration hypothesis, which states that the spatial symmetries of the system
are restored in the measurable statistical quantities.

We start our consideration with the structure functions of the following form
\begin{equation}
S_{n}(r) = \langle [\theta(t,{\bm x})-\theta(t,{\bm x'})]^{2n}\rangle.
\label{struc}
\end{equation}
Dimensionality considerations together with the RG equations give the
asymptotic expressions in the region $\mu r\gg 1$:
\begin{equation}
S_{n}(r) = (\nu\mu^{2})^{-n} (\mu r)^{-2n\Delta_{\theta}}\zeta(mr),
\label{struc2}
\end{equation}
where $r=|{\bm x'}-{\bm x}|$, $m=L^{-1}$, the critical dimensions of the fields are given by
Eq.~(\ref{ThDim}), and $\zeta$ are certain scaling functions~\cite{AAV}.

We assume that the function $C({\bm r}/L_\theta)$ in Eq.~(\ref{noise})
depends additionally on a constant unit vector ${\bm n}=\{n_i\}$ that determines
a certain distinguished direction.
Thus, the operator product expansion (OPE) in the irreducible composite operators~\cite{Zinn} 
\begin{equation}
[\theta(t,{\bm x})-\theta(t,{\bm x'})]^{2n} \simeq \sum_{F} C_{F}(mr)
F(t,{\bm x}), \quad {\bm x}=({\bm x}+{\bm x'})/2,
\label{OPE2}
\end{equation}
which is valid for $r\to 0$, provides the
expansion in the irreducible representations of the SO$(d)$ group. 
Since in order to identify all critical dimensions it is sufficient to consider unaxial anisotropy,
Eq.~(\ref{OPE2}) gives rise 
to the $d$-dimensional generalizations
of the Legendre polynomials $P_{l}(\cos\vartheta)$ (which are the basis of such representation), where $\vartheta$
is the angle variable between the vectors ${\bm r}$ and ${\bm n}$.
The structure functions~(\ref{struc}) 
are obtained by averaging~(\ref{OPE2}) with the weight
$\exp {\cal S}_{R}$, where ${\cal S}_{R}$ is the renormalized action functional~(\ref{Fact}). 
The mean values $\langle F(x)\rangle~\propto~(mr)^{\Delta_{F}}$
appear in the right hand side.

The main contribution to the ``shell'' with a given rank $l$ is determined
by the $l$th rank operator with the lowest critical dimension. 
The expansion that takes into account only the leading term in each
shell has the form 
\begin{equation}
S_{n} \simeq r^{-2n\Delta_{\theta}} \sum_{l=0}^{2n} A_{l}(mr)\,
P_{l}(\cos\vartheta)\, (mr)^{\Delta_{(2n,l)}} + \dots,
\label{shells}
\end{equation}
where we omit the dimensional factors $\nu$ and $\mu$ and the ellipsis stands
for the contributions with $l>2n$, which contain more derivatives than fields;
$A_{l}(mr)$ are the coefficient functions analytical in $mr$.
The dimensions $\Delta_{(2n,l)}$ are the critical dimensions of the operators 
\begin{equation}
F^{(n,l)}_{i_{1}\dots i_{l}} =
\partial_{i_{1}}\theta\cdots\partial_{i_{l}}\theta\,
(\partial_{i}\theta\partial_{i}\theta)^{s} + \dots,
\label{Fnp}
\end{equation}
which are constructed
solely of the gradients of the passive scalar field and have the
lowest canonical dimension (i.e., contain the minimal number of the derivatives).
Here, $l$ is the number of the free vector indices (i.e., the rank of the tensor)
and $n=l+2s$ is the total number of the fields $\theta$ entering a given
 operator. The ellipsis then represents the subtractions with Kronecker's delta
symbols that make the operator irreducible (so that the contraction with respect
to any pair of the free tensor indices vanishes). For example,
\begin{equation}
F^{(2,2)}_{ij} = \partial_{i}\theta \partial_{j}\theta -
\frac{\delta_{ij}}{d}\, (\partial_{k}\theta\partial_{k}\theta).
\label{Irr}
\end{equation}

For the pair correlation functions, the full analog of the
expression~(\ref{shells}) can be presented in the form that includes all the shells:
\begin{equation}
\langle\theta(t,{\bm x}) \theta(t,{\bm x}') \rangle = r^{-2\Delta_{\theta}}
\sum_{l=0}^{\infty} A_{l}(mr)\,
P_{l}(\cos\vartheta)\, (mr)^{\Delta_{l}}.
\label{Orujo}
\end{equation}
This is a consequence of the expression 
\begin{equation}
F(x) \partial G(x) = - G(x) \partial F(x) + \partial [F(x) G(x)]
\label{Nwt}
\end{equation}
for the operators $F(x)$ and $G(x)$ of the form
\begin{equation}
F_{i_{1}\dots i_{l}}(x) = \theta(x) \partial_{i_{1}}
\cdots\partial_{i_{l}} \theta(x) + \dots,
\label{Fl}
\end{equation}
where the ellipsis stands for the subtractions with 
Kronecker's delta symbols that make the operator irreducible. 
It is clear that for the pair correlation function the leading term of the $l$th shell 
is determined by the single operator~(\ref{Fl}) with two fields $\theta$ and $l$ tensor indices. 
From Eq.~(\ref{Nwt}) it follows that this operator is unique up to
derivatives, which have vanishing mean values and do not contribute to the
quantities of interest; see Secs.~IVC and VC in Ref.~\cite{AK14} for detailed discussion. The 
critical dimensions $\Delta_l$ in Eq.~(\ref{Orujo}) are dimensions of the composite operators~(\ref{Fl}).

Furthermore, from the relation~(\ref{Nwt}) it follows that for odd $l$ the operator~(\ref{Fl})
itself reduces to a linear combination of the derivatives. In the following, we will be interested
only in the operators which are not reducible to derivatives and will consider only even values of $l$. 

\section{Inertial range asymptotic behavior of the pair correlation function}
\label{sec:final}

In general, a local composite
operator is a polynomial constructed from the primary fields
$\Phi(x)$ and their finite-order derivatives at a single space-time point
$x=\{t,{\bm x}\}$. Due to a coincidence of the field arguments, new UV
divergences arise in the Green's functions with such objects~\cite{Zinn,Vasiliev}.
The total canonical dimension of an arbitrary 1-irreducible Green's function
$\Gamma = \langle F\,\Phi\dots\Phi\rangle$
that includes one composite operator $F$ and an arbitrary number of the 
primary fields $\Phi$ (the formal index of UV divergence) is given by the relation
\begin{equation}
\label{D-Comp}
d_{\Gamma}=d_F-\sum_\Phi N_\Phi d_\Phi,
\end{equation}
where $N_{\Phi}$ is the number of the fields $\Phi$ {entering $\Gamma$},
$d_{\Phi}$ is the total canonical dimension of the given field $\Phi$, $d_{F}$ is the canonical
dimension of the operator.
In the process of the renormalization operators can mix with each other,
\begin{equation}
F_{i}=\sum _{j} Z_{ij} F_{j}^{R},
\end{equation}
where $Z_{ij}$ is the renormalization matrix. 

We are interested in the scaling dimensions of the operators~(\ref{Fl}), which are given by
the eigenvalues of the matrix~$\Delta_{F}$ [see~(\ref{Krit})] calculated for the mixed operators.
Since the original stochastic equation~(\ref{density1}) is linear in field
$\theta$, the necessary diagrams for a calculation of the matrix $Z_{ij}$ do not contain the
propagator $\langle\theta\theta\rangle_{0}$ from Eq.~(\ref{lines3}). Hence, all the calculations can be 
performed directly in
the model without the random noise in Eq.~(\ref{density1}), i.e., in the SO$(d)$ covariant model, where the 
irreducible tensor operators with different ranks cannot mix in renormalization procedure. 
The only possibility to mix during the renormalization is mixing within the operator's own ``family'' of derivatives:
the operators with additional derivatives or with the fields $\theta'$ and $v_i'$ have
too high canonical dimensions, the appearance of a field $v_i$ is forbidden
by the (restored) Galilean symmetry, and additional fields $\theta$ are forbidden by the linearity of the model.
Herewith, relation~(\ref{Nwt}) shows that all the other operators obtained the same rank 
differ from~(\ref{Fl}) by a total derivative and, therefore, give the same contribution into the OPE.
This means that the matrix $\Delta_{F}$ is in fact
triangular and the composite operators~(\ref{Fl}) can be treated as multiplicatively renormalizable,
$F^{(n,l)} = Z_{(n,l)} F^{(n,l)}_{R}$, with certain renormalization
constants $Z_{(n,l)}$ denoted later for simplicity as $Z_l$.

Let us introduce $\Gamma_n(x;\theta)$, the $\theta^{n}$ term of the
expansion in $\theta(x)$ of the generating functional of 
1-irreducible Green's functions with one composite operator $F(x)$
and any number of the fields $\theta$:
\begin{eqnarray}
\Gamma_{n}(x;\theta) = \int d x_{1} \cdots \int d x_{n}
\langle F(x) \theta(x_{1})\cdots\theta(x_{n})\rangle \times \theta(x_{1})\cdots\theta(x_{n}).
\label{Gamma1}
\end{eqnarray}
The renormalization constants $Z_l$ are determined by the requirement that the {1-irreducible} 
functions~(\ref{Gamma1})
are UV finite in the renormalized theory.

\begin{figure}[t]
  \center
    \includegraphics[width=.18\textwidth,clip]{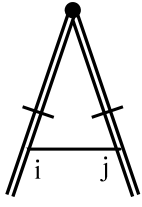}
  \caption{The one-loop approximation of the function $\Gamma_{n}(x;\theta)$. 
Latin letters $i$ and $j$ denote the internal indices of the velocity field with implied summation.}
\label{fig:Comp}
\end{figure}

The one-loop approximation for the 1-irreducible function $\Gamma_{n}(x;\theta)$ can be formally written as
\begin{equation}
\Gamma_{n}(x;\theta)= F(x) + \frac{1}{2} \widetilde{\Delta},
\label{Gamma2}
\end{equation}
where the first term is the tree (loop-less) approximation, $\widetilde{\Delta}$ is
the one-loop graph depicted in Fig.~\ref{fig:Comp}, and $1/2$ is the symmetry coefficient of the given graph.
The dot with two attached lines in the top of the diagram denotes the operator vertex, i.e.,
the variational derivative
\begin{equation}
V(x;x_{1},x_{2})=\delta^{2}F(x)/{\delta\theta(x_{1})\delta\theta(x_{2})}.
\label{Vae}
\end{equation}
The contribution of a specific diagram into the functional~(\ref{Gamma2})
for any composite operator $F$ is represented in
the form
\begin{equation}
\label{Diag-General}
\Gamma_{n} = V\times I\times
\theta \dots \theta,
\end{equation}
where $V$ is the vertex factor given by Eq.~(\ref{Vae}),
$I$ is the diagram itself, and the product $\theta \dots \theta$ corresponds
to the external tails.

The calculation of the renormalization constant $Z_{l}$ and anomalous dimension $\gamma_{l}$ 
(see Appendix~\ref{a3} for details) gives
\begin{eqnarray}
\label{Zl}
Z_{l}&=& 1 - \frac{g}{2(w^2+1)}\frac{1}{y}\, {\cal S}_{l}(d),\\
\label{gaml}
\gamma_{l}&=& \frac{g}{2(w^2+1)}\, {\cal S}_{l}(d).
\end{eqnarray}
The factor ${\cal S}_{l}(d)$ in Eqs.~(\ref{Zl}) and~(\ref{gaml}) denotes the double sum
\begin{eqnarray}
{\cal S}_{l}(d) = \sum_{s,m=0}^{s+m+2\le l}
\frac{(-1)^{s+m}2^{s} C_{l}^{s+m+2}(s+m+2)!}
{d(d+2)\dots [d+2(s+m)+2]}\left(\frac{2w^2}{w^2+1}\right)^m.
\label{Sl}
\end{eqnarray}
This sum can be calculated for any given $l$:
\begin{eqnarray}
{\cal S}_{l}(d)=\frac{l(l-1)}{4z}
\left[\frac{w^2+1}{l+z-1} - \frac{w^2}{z+1}\, {}_2F_1\left(1,2-l;2+z;\frac{w^2}{w^2+1}\right)\right],
\end{eqnarray}
where $z = d/2$ and ${}_2F_1$ is the hypergeomeric function, defined for $|t|<1$ as
\begin{equation}
{}_2F_1(a, b; c; t) = 1 + \sum_{k=1}^{\infty}\left[\prod_{l=0}^{k-1} \frac{(a+l)(b+l)}{(1+l)(c+l)}\right]\, t^k;
\end{equation}
see Appendix~\ref{b} for details.

\begin{figure}[t]
   \centering
    \begin{tabular}{c}
     \includegraphics[width=9.0cm]{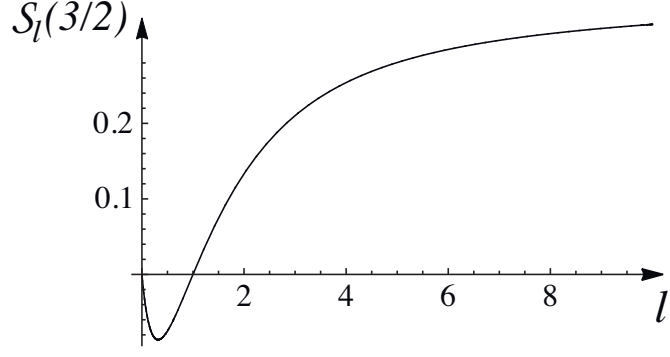}
    \end{tabular}
   \caption{The graph of the function ${\cal S}_{l}(3/2)$ as a function of $l$.}
  \label{fig:graph}
\end{figure}

The critical dimension of the operator~(\ref{Fl}) is obtained using the relations~(\ref{Krit}) 
and~(\ref{ThDim})
\begin{eqnarray}
\Delta_{l} = l+2\Delta_{\theta}+\gamma_{l}^{*}= l-2+y/3+ \gamma_{l}^{*},
\label{Dl}
\end{eqnarray}
where $\gamma_{l}^{*}$ is the value of $\gamma_{l}$ at the fixed point. 
For the realistic values of the parameters $d=3$ and $w^*\approx1.393$, see Eq.~(\ref{FP}), we
get
\begin{eqnarray}
\gamma_{l}^{*}\approx 0.566\,{\cal S}_l(3/2) \times y
\label{gamZ}
\end{eqnarray}
with the higher-order corrections in scaling exponent $y$. The factor ${\cal S}_{l}(3/2)$ is
given by the expression
\begin{equation}
{\cal S}_{l}(3/2)=\frac{l(l-1)}{6}\, \left[\frac{2.94}{l+1/2} - 0.776\, {}_2F_1\left(1,2-l;7/2;0.66\right)\right]
\end{equation}
and depends on $l$ as depicted in Fig.~\ref{fig:graph}; the same type of behavior is 
valid for any spatial dimension $d\geq3$.

It is important that ${\cal S}_{l}(3/2)>0$ for $l\geq1$: 
this leads to a monotonical increase of $\Delta_l$ as $l\to\infty$, see 
expression~\eqref{Dl}. Moreover, $\Delta_0<0$.
The quantity of the spatial derivatives $l$ illustrate the degree of anisotropy: the larger $l$ 
the higher the degree of the anisotropy, see Eq.~(\ref{Orujo}). 
Thus, there is a hierarchy of the anisotropic contributions in the inertial range 
asymptotic behavior of the pair correlation function~(\ref{Orujo}), and
the leading term is given by the scalar operator~$\theta^{2}$.

This fact has a clear physical interpretation: in the inertial range, the
leading contribution is given by the isotropic shell $l = 0$
and coincides with the scalar isotropic model, 
while the terms with $l \geq 2$ provide only corrections which become insignificant as $mr \to 0$.
Moreover, the corrections become less pronounced with increasing $l$, i.e., with 
increasing degree of the anisotropy. This effect confirms Kolmogorov's hypothesis of the local
isotropy restoration.

It is worth mentioning that same analogy exists between the 
present hierarchy and the well-known multipole expansion in ordinary classical electrostatics which 
can also be written as an expansion in spherical harmonics~\cite{Jackson}. 
The multipole expansion can be viewed as a series with progressively finer angular features. The
initial isotropic term, corresponding to the potential of a point-like charge, gives the leading contribution at large distances. The other terms are anisotropic (they involve angular dependence) and give corrections that decay faster and faster as the ``degree of anisotropy'' $l$ increases.

This is not just a superficial analogy. It becomes especially clear if one applies the zero-mode approach to the
advection problem in the Kraicnan's rapid-change model (or in some analogous model, in which turbulent flow is
simulated by some Gaussian statistics); see, e.g., Ref.~\cite{LM}. 
Employing the zero-mode approach terminology, the individual terms
of the spherical harmonics expansion in electrostatics are the homogeneous solutions (the so-called zero modes) of the Poisson
equation. If we are interested in the asymptotic behavior at small or large distances, only the zero modes restricted
at the origin or at infinity should be taken into account, respectively. 
The difference with electrostatic problems is that for the
turbulent advection the differential operator is more complicated. Moreover, since we are 
interested in the inertial
range behavior (i.e., behavior in the interval, restricted by both large and small scales) it is not so
simple to choose
the right solution from two possible zero modes. Nevertheless, the result is similar to the electrostatic
case: the leading
term corresponds to $l = 0$ and is isotropic, while the other (anisotropic) contributions provide the 
decaying corrections and
obey the hierarchy with respect to the value of $l$.

\section{Conclusion}
\label{sec:concl}

In this paper incompressible fluid is
studied using the field theoretic approach. We have considered the stochastic Navier-Stokes equation with
colored noise (i.e., the model with arbitrary finite correlation time of the velocity field) to
describe fluid dynamics. The second problem considered is the 
advection of the passive scalar field by this turbulent flow. The critical dimensions of the fields are
calculated for both problems.

Within the one-loop approximation the only nontrivial regime of the long-distance (IR) behavior is found to be reduced to the limiting case of the rapid-change type behavior.
This regime is realized for $y>0$, $\eta>y/3$, where 
$y$ and $\eta$ describe the energy spectrum ${\cal E} \propto k^{1-y}$ and the dispersion
law $\omega \sim k^{2-\eta}$ of the velocity field.
The second nontrivial fixed point, existing if 
$\frac{1}{3}<\frac{y}{\eta}<\frac{1}{3}+\frac{4/3}{3d(d-1)+2}$, 
is a saddle type point. 
The fact that the only nontrivial IR attractive fixed point corresponds to vanishing correlation time
means, in particular, that the Galilean symmetry, violated by the 
colored stochastic force, is automatically restored in the IR limit. As it should be for the case of rapid-change behavior, the calculated critical dimensions of the fields coincide with the results obtained for the zero-time model which was considered earlier in Ref.~\cite{NS-Zero}. 

The inertial-range behavior of the correlation function of two composite operators constructed from the
advected fields $\theta$ was studied using the OPE. Existence of the anomalous scaling (singular 
power-like dependence on the integral scale $L$) was established. From the leading-order (one-loop) 
calculations it follows that the main contribution into the OPE is given by the isotropic term 
corresponding to $l=0$, where $l$ is the number of the Legendre polynomial entering the expansion of
the correlation function and signifies a degree of the anisotropy; all other terms with $l\geq2$ provide 
only corrections.

These two facts (the restoration of the Galilean symmetry and isotropy restoration)
give a quantitative illustration of the general concept that the symmetries of the Navier--Stokes 
equation, broken spontaneously and by initial or boundary conditions, are restored in the statistical 
sense for fully developed turbulence~\cite{Legacy,LegacyD,Monin,K41a,K41b,K41c}. 


\section*{Acknowledgments}

The authors are indebted to Mikhail V. Kompaniets and Igor Altsybeev for discussions and to Tom\'a\v{s} Lu\v{c}ivjansk\'y for critical reading of the manuscript.
N.M.G. acknowledges the support from the Saint Petersburg Committee of Science and High School, M.M.K. was supported by the Basis Foundation.


\appendix

\section{Calculation details for the model of incompressible fluid}
\label{calc}

This section contains detailed calculations of the diagrams, defining the renormalization
constants $Z_1$ and $Z_2$ (see Sec.~\ref{RenModel}).
All calculations are performed in the analytical regularization and MS scheme.

\subsection{Calculation of the self-energy operator}
\label{a1}

Let us start with the graph presented in Fig.~\ref{fig:SelfEnergy}. 
The analytical expression for it reads
\begin{eqnarray}
\label{Sigma-Expr}
\Sigma_{\alpha\beta}&=&D_0\int\frac{d\omega}{2\pi}\int\frac{d{\bm k}}
{(2\pi)^d} V_{\alpha ab}({\bm p})V_{dc\beta}({\bm p+\bm k})P_{ac}({\bm p+\bm k})P_{bd}({\bm k}) \nonumber \\
&\times&\frac{k^{8-d-y-2\eta}}{(\omega^2+\nu_0^2 u_0^2 k^{4-2\eta})(\omega^2+\nu_0^2k^4)(-i\omega+\nu_0|{\bm p+\bm k}|^2)}.
\end{eqnarray}
Here and below $V_{ijk}({\bm p})$ is the triple vertex~(\ref{Triple-vertex});
Greek letters $\alpha$, $\beta$ and Latin letters $a$~-- $d$
denote the vector indices of the propagators~(\ref{VV})
and~(\ref{VVprime}) with the implied summation
over the repeated indices. Since the index of divergence for this diagram
$d_\Gamma=2$, we have to calculate only the terms proportional
to $p^2$, where $\bm p$ denotes an external momentum.

Let us consider the expression~(\ref{Dyson}).
Taking trace of both sides of it one obtains the scalar equation
\begin{equation}
\Gamma_2=i\omega -\nu_0 p^2+\Sigma, 
\end{equation}
where we have introduced for brevity
\begin{equation}
\label{Sigma-Sc}
\Sigma = \Sigma_{\alpha\beta}P_{\alpha\beta}/(d-1).
\end{equation}
The calculation of the (scalar) index structure $J$ of the quantity $\Sigma$ yields
\begin{eqnarray}
\label{J}
J&=&V_{\alpha ab}({\bm p})V_{dc\beta}({\bm p+\bm k})P_{ac}({\bm p+\bm k})P_{bd}({\bm k})P_{\alpha \beta}({\bm p}) \\ \nonumber
&\cong& ({\bm A\cdot\bm p}) + B p^2 + {\it O}({p}^3).
\end{eqnarray}
The vector coefficient ${\bm A}$ and scalar coefficient $B$ are
\begin{eqnarray}
\label{Alpha}
{\bm A} &=& 2{\bm k}, \\
\label{Beta}
B&=& \sin^4\varphi-3\sin^2\varphi\cos^2\varphi+(2-d)\sin^2\varphi,
\end{eqnarray}
where $\varphi$ is the angle between the external momenta ${\bm p}$ and internal momenta ${\bm k}$.

The integration over the frequency $\omega$ in the expression~(\ref{Sigma-Expr}) gives 
\begin{eqnarray}
\label{Int-omega}
\int\frac{d\omega}{2\pi}
\frac{1}{(\omega^2+\nu_0^2 u_0^2 k^{4-2\eta})(\omega^2+\nu_0^2k^4)(-i\omega+\nu_0|{\bm p+\bm k}|^2)} \nonumber \\
=\frac{1}{2\nu_0^4}\frac{u_0k^{2-\eta}+k^2+|{\bm p+\bm k}|^2}{u_0k^2k^{2-\eta}(k^2+u_0k^{2-\eta})(u_0k^{2-\eta}+|{\bm p+\bm k}|^2)(k^{2}+|{\bm p+\bm k}|^2)}.
\end{eqnarray}
Combining the expression~(\ref{J}) with the numerator of the expression~(\ref{Int-omega}) gives
\begin{equation}
\label{Sigma-AB}
\Sigma\propto 
\int\frac{d{\bm k}}{(2\pi)^d} 
\frac{\widehat{{\bm A}}+\widehat{{\bm B}}}{u_0k^{4-\eta}(k^2+u_0k^{2-\eta})(u_0k^{2-\eta}+|{\bm p+\bm k}|^2)(k^{2}+|{\bm p+\bm k}|^2)}.
\end{equation}
Here, the coefficients $\widehat{{\bm A}}$ and $\widehat{{\bm B}}$ denote the 
quantities proportional to $p^2$ and $\bm p$, respectively:
\begin{eqnarray}
\label{A}
\widehat{{\bm A}} &=& p^2 (u_0k^{2-\eta}+2k^2){\bm A}+2B ({\bm p\cdot\bm k}){\bm p}, \\ 
\label{B}
\widehat{{\bm B}} &=& B (u_0k^{2-\eta}+2k^2){\bm p},
\end{eqnarray}
with the vector ${\bm A}$ and scalar $B$ defined in the expressions~(\ref{Alpha}) and~(\ref{Beta}).

The integration over the internal momenta ${\bm k}$ can be
simplified in the MS scheme, in which all the anomalous
dimensions $\gamma_{1,2}$ are independent of the regularizers like $y$
and $\eta$. Hence, we may choose them arbitrarily with the only restriction that
our diagrams have to remain UV finite; see Ref.~\cite{FinTimeEta} for
detailed discussion. The most convenient way is to put $\eta=0$, so
expanding the denominator of~(\ref{Sigma-AB}), combining it with~(\ref{A})~-- (\ref{B}) and taking
into account that we are interested only in the term proportional to ${\bm p}$ for $\widehat{{\bm B}}$ and
to $p^0$ for $\widehat{{\bm A}}$ one obtains
\begin{eqnarray}
\label{Sigma-K}
\Sigma_{\alpha\beta}\propto 
p^2\frac{1}{2(u_0+1)}\int\frac{d{\bm k}}{(2\pi)^d} 
\frac{1}{k^2}\biggl[A(u_0+2)-4\sin^2\varphi\cos^2\varphi 
+\frac{2(u_0+2)(u_0+3)}{u_0+1}\sin^2\varphi\cos^2\varphi\biggr];
\end{eqnarray}
the coefficient $A$ is defined in Eq.~(\ref{Alpha}).

In order to
integrate over the vector ${\bm k}$ we need to
average the expression~(\ref{Sigma-K}) over the angles:
\begin{equation}
\label{Int-Aver-Angles}
\int d{\bm k} f({\bm k})=
S_{d} \int_m^\infty dk\,k^{d-1}\,
\left\langle f({\bm k})\right\rangle,
\end{equation}
where $\langle\cdots\rangle$ is the averaging over the unit sphere in the
$d$-dimensional space, $S_{d}$ is its surface area, and
$k= |{\bm k}|$. To average the function~(\ref{Sigma-K}) over
the angles in the orthogonal subspace we use the relations
\begin{eqnarray}
\label{Cos-Angles}
\left\langle \cos^2\varphi\right\rangle &=& \frac{1}{d}, \\ \nonumber
\left\langle \cos^4\varphi\right\rangle &=& \frac{3}{d(d+2)}.
\end{eqnarray}
This gives
\begin{eqnarray}
\label{Sigma-X}
\Sigma\propto 
p^2\frac{2}{(u_0+1)^2}\int\frac{d{\bm k}}{(2\pi)^d} 
\frac{1}{k^2} X,
\end{eqnarray}
where 
\begin{eqnarray}
\label{X}
X = \frac{d-1}{d(d+2)}\left[u_0^2d(d-1)+3u_0d(d-1)+2d^2-2d+4\right].
\end{eqnarray}
Combining the expression~(\ref{Sigma-Expr}) with the expressions~(\ref{Sigma-Sc}), 
(\ref{Int-omega}), (\ref{Sigma-X}), and expression~(\ref{D0}) for the amplitude $D_0$ we find
\begin{eqnarray}
\label{Sigma-Vec}
\Sigma_{\alpha\beta} =
-\frac{1}{4}p^2g_0\nu_0\frac{u_0}{(u_0+1)^3}X
\int_m^\infty\frac{dk}{(2\pi)^d} \frac{1}{k^{d+y}}.
\end{eqnarray}
Taking into account the multiplier $(d-1)^{-1}$ coming from~(\ref{Sigma-Sc}), after the 
integration of the expression~(\ref{Sigma-Vec}) over the modulus $k$ one obtains
\begin{equation}
\label{Sigma}
\Sigma =
-\frac{1}{4}p^2g_0\nu_0\frac{u_0^3d(d-1)+3u_0^2d(d-1)+2u_0(d^2-d+2)}{d(d+2)(u_0+1)^3}
C_d\frac{m^{-y}}{y},
\end{equation}
where $C_{d}\equiv S_{d}/(2\pi)^{d}$.
Combining this expression with Eq.~(\ref{Dyson}) from Sec.~\ref{RenModel} one immediately obtains
the renormalization constant $Z_2=Z_\nu$.

\subsection{Calculation of the vertex diagrams}
\label{a2}

Let us start with the graph presented in Fig.~\ref{fig:Trio}a.
The analytical expression for it is 
\begin{eqnarray}
\label{D1}
\Delta_1&=&D_0\int\frac{d\omega}{2\pi}\int\frac{d{\bm k}}
{(2\pi)^d} V_{\alpha cd}({\bm p})V_{a\beta b}(-\bm k)V_{ef\gamma}({\bm p+\bm k}) \\
&\times&\frac{P_{ac}(\bm k)P_{de}({\bm p +\bm k})P_{bf}({\bm q-\bm k })}{(i\omega+\nu_0 k^2)(-i\omega+\nu_0|{\bm p+\bm k}|^2)}
\frac{(q-k)^{8-d-y-2\eta}}{(\omega^2+\nu_0^2|\bm q-\bm k|^4)(\omega^2+\nu_0^2 u_0^2 |\bm q-\bm k|^{4-2\eta})}.\nonumber 
\end{eqnarray}
Here and below $\bm p$ and $\bm q$ are the external momenta, $\bm k$ denotes an internal (loop) momenta;
the index of divergence for this diagram $d_\Gamma=1$, therefore, we need to calculate only the terms,
proportional
to ${\bm p}^1$ or ${\bm q}^1$. Since $\Delta_1\propto V_{\alpha cd}({\bm p})$, we may set $q=p=0$ in all
the other multipliers. This observation significantly simplifies
the expression for the divergent part
of Eq.~(\ref{D1})
\begin{eqnarray}
\label{D1s}
\Delta_1&\cong&D_0\int\frac{d\omega}{2\pi}\int\frac{d{\bm k}}
{(2\pi)^d} V_{\alpha cd}({\bm p})V_{a\beta b}(-\bm k)V_{ef\gamma}({\bm k}) \\
&\times&\frac{P_{ac}(\bm k)P_{de}({\bm k})P_{bf}({\bm k })}{(i\omega+\nu_0{ k}^2)(-i\omega+\nu_0 k^2)}\frac{k^{8-d-y-2\eta}}{(\omega^2+\nu_0^2k^4)(\omega^2+\nu_0^2 u_0^2 k^{4-2\eta})}.\nonumber 
\end{eqnarray}
Integration over the frequency at $\eta=0$ leads to
\begin{eqnarray}
\label{D1-res}
\int\frac{d\omega}{2\pi}\frac{1}{(i\omega+\nu_0{ k}^2)(-i\omega+\nu_0 k^2)(\omega^2+\nu_0^2k^4)(\omega^2+\nu_0^2 u_0^2 k^{4-2\eta})}
=\frac{u_0+2}{4k^{10}\nu_0^5u_0(u_0+1)^2}.
\end{eqnarray}
The calculation of the index structure $J^1_{\alpha\beta\gamma}$ of the quantity $\Delta_1$ 
is straighforward
\begin{eqnarray}
\label{tmp}
J^1_{\alpha\beta\gamma}&=&V_{\alpha cd}({\bm p})V_{a\beta b}(-\bm k)V_{ef\gamma}({\bm k})P_{ac}(\bm k)P_{de}({\bm k})P_{bf}({\bm k })\\ \nonumber
&=&2i\left[p_\alpha-\frac{(\bm p\cdot \bm k)k_\alpha}{k^2}\right]k_\beta k_\gamma.
\end{eqnarray}
In order to integrate the expression~(\ref{tmp}) over the vector $\bm k$ we employ
the relations similar to Eqs.~(\ref{Cos-Angles})
\begin{eqnarray}
\label{k-Angles}
\left\langle \frac{k_i k_j}{k^2} \right\rangle
&=& \frac{\delta_{ij}}{d}, \label{k-Angles0}\\ \nonumber
\left\langle \frac{k_i k_jk_lk_m}{k^4} \right\rangle
&=& \frac{\delta_{ij}\delta_{lm}+\delta_{il}\delta_{jm}+\delta_{im}\delta_{jl}}{d(d+2)}.
\end{eqnarray}
Taking into account the expressions~(\ref{D1-res})~-- (\ref{k-Angles}), for the divergent part of
the diagram $\Delta_1$ one obtains
\begin{equation}
\label{D1Ans}
\Delta_1=i\,g_0\frac{u_0(u_0+2)}{2(u_0+1)^2}\frac{(d+1)p_\alpha\delta_{\beta\gamma}-p_\beta\delta_{\alpha\gamma}-p_\gamma\delta_{\alpha\beta}}{d(d+2)}C_d\frac{m^{-y}}{y}.
\end{equation}
\newline

The analytical expression for the graph presented in Fig.~\ref{fig:Trio}b is
\begin{eqnarray}
\label{D2}
\Delta_2&=&D_0\int\frac{d\omega}{2\pi}\int\frac{d{\bm k}}
{(2\pi)^d} V_{\alpha cd}({\bm p})V_{a\beta b}(\bm p-\bm k)V_{fe\gamma}({\bm q-\bm k}) \\
&\times&\frac{P_{ac}(\bm p-\bm k)P_{de}({\bm k})P_{bf}({\bm q-\bm k })}{(i\omega+\nu_0|\bm p-{\bm k}|^2)(i\omega+\nu_0|{\bm q-\bm k}|^2)}\frac{k^{8-d-y-2\eta}}{(\omega^2+\nu_0^2k^4)(\omega^2+\nu_0^2 u_0^2 k^{4-2\eta})}.\nonumber 
\end{eqnarray}
Like in the previous case, since $\Delta_2\propto V_{\alpha cd}({\bm p})$ we may set $q=p=0$ in all
the other multipliers: 
\begin{eqnarray}
\label{D2s}
\Delta_2&\cong&D_0\int\frac{d\omega}{2\pi}\int\frac{d{\bm k}}
{(2\pi)^d} V_{\alpha cd}({\bm p})V_{a\beta b}(\bm k)V_{fe\gamma}({\bm k}) \\
&\times&\frac{P_{ac}(\bm k)P_{de}({\bm k})P_{bf}({\bm k })}{(i\omega+\nu_0k^2)^2}\frac{k^{8-d-y-2\eta}}{(\omega^2+\nu_0^2k^4)(\omega^2+\nu_0^2 u_0^2 k^{4-2\eta})}.\nonumber 
\end{eqnarray}
The integration over the frequency $\omega$ at $\eta=0$ leads to the expression 
\begin{eqnarray}
\label{D2-res}
\int\frac{d\omega}{2\pi}\frac{1}{(i\omega+\nu_0{ k}^2)^2(\omega^2+\nu_0^2k^4)(\omega^2+\nu_0^2 u_0^2 k^{4-2\eta})}=\frac{u_0(u_0+3)+4}{8k^{10}\nu_0^5u_0(u_0+1)^3}.
\end{eqnarray}
The calculation of the index structure $J^2_{\alpha\beta\gamma}$ gives 
\begin{eqnarray}
\label{tmp2}
J^2_{\alpha\beta\gamma}&=&V_{\alpha cd}({\bm p})V_{a\beta b}(\bm k)V_{fe\gamma}({\bm k})P_{ac}(\bm k)P_{de}({\bm k})P_{bf}({\bm k })\\ \nonumber
&=&-2i p_cP_{c\alpha}(\bm k) k_\beta k_\gamma.
\end{eqnarray}
Taking into account the expressions~(\ref{D2-res}) and~(\ref{tmp2}), the integration of the expression~(\ref{D2s}) over the momenta $\bm k$ gives
\begin{equation}
\label{D2Ans}
\Delta_2=-i\,g_0\frac{u_0(u_0^2+3u_0+4)}{4(u_0+1)^3}\frac{(d+1)p_\alpha\delta_{\beta\gamma}-p_\beta\delta_{\alpha\gamma}-p_\gamma\delta_{\alpha\beta}}{d(d+2)}C_d\frac{m^{-y}}{y}.
\end{equation}
\newline

The analytical expression for the divergent part of the graph presented in Fig.~\ref{fig:Trio}c coincides
with the expression~(\ref{D2s}). Thus, $\Delta_3=\Delta_2$ and it is also given by the expression~(\ref{D2Ans}).

Using the transversality condition $\partial_i v_i=0$ together with the
expression~(\ref{Triple-vertex}) and moving the derivative in the vertex 
from the field ${\bm v}'$ onto the field ${\bm v}$ we conclude that the term proportional
to a momentum $p_\alpha$ gives no contribution, therefore for the sum of the three
triangle diagrams $\Delta_1$, $\Delta_2$, and $\Delta_3$ [see Eqs.~(\ref{D1Ans}) and~(\ref{D2Ans})] one
obtains 
\begin{equation}
\label{VVV-Ans}
\Delta_1+\Delta_2+\Delta_3=i\,g_0\frac{u_0}{(u_0+1)^3}\frac{p_\beta\delta_{\alpha\gamma}+p_\gamma\delta_{\alpha\beta}}{d(d+2)}C_d\frac{m^{-y}}{y}.
\end{equation}
Combining this expression with Eq.~(\ref{VVV}) from Sec.~\ref{RenModel} one immediately obtains the renormalization constant $Z_1=Z_v$.

\section{Calculation details for the model of passive advection}
\label{AppAdd}

This section contains detailed calculations of the diagram, defining the renormalization constant
$Z_l$, and the
double sum ${\cal S}_{l}(d)$, entering in the expression for the anomalous dimension $\gamma_l$ 
(see Sec.~\ref{sec:final}).
The calculation of renormalization constant is performed in the analytical regularization and the MS scheme.

\subsection{Calculation of the diagram with insertion of the composite operator}
\label{a3}

The only graph $\widetilde{\Delta}$ which is
required for the critical dimensions of the correlation functions~(\ref{Orujo}) is presented in Fig.~\ref{fig:Comp}.
To simplify the process of calculations it is convenient to contract the operator~(\ref{Fl}) with a
constant vector ${\bm \lambda}=\{\lambda_{i}\}$.
As a result one obtains the scalar operator
\begin{equation}
F_{l} = \theta (\lambda_{i}\partial_{i})^{l} \theta + \dots,
\label{FlSk}
\end{equation}
where the terms, denoted by the ellipsis, necessarily involve the
factors of $\lambda^{2}$. The appearance of $\lambda^{2}$ means that the corresponding initial operator contains 
$\partial^2$, i.e., its canonical dimension is too high. Therefore, we should omit the terms with $\lambda^{2}$.
The vertex factor~(\ref{Vae}) in this case takes the form
\begin{equation}
V(x;x_{1},x_{2}) = \delta(x-x_{1}) (\lambda_{i}\partial_{i})^{l}
\delta(x-x_{2}) + \{ x_{1} \leftrightarrow x_{2} \}.
\label{Wae}
\end{equation}

Let us choose the external momentum ${\bm p}$ to flow into the diagram
through the left lower vertex and to flow out of the diagram through the right lower one.
Since the divergence of the graph $\widetilde{\Delta}$ is logarithmic, the external momentum
flowing through the operator vertex and all the
external frequencies are set equal to zero. 
We are interested in the
value of the anomalous dimension at the fixed point, therefore, we may perform the 
substitutions $g=g^*$, $w=w^*$ [see Eq.~(\ref{FP})] and $u\to\infty$ from the very
beginning; furthermore, the limiting case $u\to\infty$ means that $(u_0\nu_0)^2\gg\omega^2$, so
the propagator function~(\ref{VV}) reads
\begin{equation}
\label{VV-red}
\left\langle v_iv_{j}\right\rangle_0 =
g_0\nu_0^3\
\frac{k^{4-d-y}}{\omega^2+\nu_0^2k^4}P_{ij}({\bm k}).
\end{equation}
Thus, the core of the diagram takes the form
\begin{equation}
\widetilde{\Delta}\cong p_{i}p_{j}\, \int \frac{d\omega}{2\pi} \int_{k>m}
\frac{d{\bm k}}{(2\pi)^{d}} \,
2i^{l} ({\bm \lambda}\cdot{\bm q})^{l} \,
\frac{g\mu^y \nu^3 k^{4-d-y}}{\omega^{2}+ \nu^2 k^{4}}\,
P_{ij} ({\bm k}) \,
\frac{1}{\omega^{2}+ w^2 \nu^2 q^{4}}.
\label{Hard}
\end{equation}
Here the factor $p_{i}p_{j}$ comes from the vertices $V_j$ [see Eq.~(\ref{vertex1})] and from the observation
that in the case of incompressible carrying fluid 
the derivative can act directly on the field $\theta$;
the factor $2i^{l}({\bm \lambda} \cdot{\bm q})^{l}$ comes from
the vertex~(\ref{Wae}) for even $l$ (for odd $l$ the two terms in
Eq.~(\ref{Wae}) would cancel each other), the factors depending on ${\bm k}$ represent
the velocity correlation function~(\ref{VV-red}), the last factor comes from the propagators 
$\langle\theta'\theta\rangle_{0}$;
the momentum ${\bm k}$ flows through the velocity propagator, so that
${\bm q}= {\bm k}+{\bm p}$. 

In order to find the corresponding
renormalization constant it is sufficient to
retain in the result for the counterterm only the terms of the same form, i.e.,
the terms of the form $({\bm \lambda} \cdot{\bm p})^{l}$, and drop all the other terms
containing $\lambda^{2}$ or $p^{2}$. 
Thus, the structure with $P_{ij} ({\bm k})$ is simplified and takes the form
\begin{equation}
p_ip_{j} P_{ij}({\bm k})\cong -({\bm p\cdot \bm k})^{2} /k^{2}. 
\end{equation}
The integration over the frequency $\omega$ in Eq.~(\ref{Hard}) is straightforward 
\begin{equation}
\widetilde{\Delta}\cong -g \mu^y \, i^{l}
\int_{k>m} \frac{d{\bm k}}{(2\pi)^{d}} \,({\bm p\cdot \bm k})^{2}
(\bm \lambda \cdot {\bm q})^{l} \,
\frac{k^{-d-y}}{q^{2}(k^{2}+w^2 q^{2})}.
\label{ioo}
\end{equation}
Expanding all the denominators in the integrand of Eq.~(\ref{ioo})
in ${\bm p}$ together with dropping all the terms with $p^{2}$ gives
\begin{eqnarray}
\label{iooooo}
\frac{1}{q^{2}} \simeq \frac{1}{k^{2}+2({\bm p\cdot \bm k})} = \frac{1}{k^{2}}
\sum^{\infty}_{s=0} \frac{(-2)^{s}({\bm p\cdot \bm k})^{s}}{k^{2s}};
\end{eqnarray}
\begin{eqnarray}
\frac{1}{k^{2}+w^2 q^{2}} &\simeq & \frac{1}{k^{2}(w^2+1)+2w^2 ({\bm p\cdot \bm k})} 
= \frac{1}{(w^2+1) k^2} \sum_{m=0}^{\infty} \frac{(-1)^{m}({\bm p\cdot \bm k})^{m}}{k^{2m}}\left(\frac{2w^2}{w^2+1}\right)^m.
\label{Tay}
\end{eqnarray}
Expanding the numerator of~(\ref{ioo}) using Newton's binomial formula
(note, that ${\bm q}= {\bm k}+{\bm p}$) one obtains
\begin{eqnarray}
({\bm \lambda}\cdot {\bm q})^{l} = \sum_{n=0}^{l}
C_{l}^{n} ({\bm \lambda}\cdot {\bm k})^{n}
({\bm \lambda}\cdot {\bm p})^{l-n}.
\label{Bin}
\end{eqnarray}
Combining the expressions~(\ref{iooooo})~-- (\ref{Bin}) one obtains threefold series over $n,m$, and $s$:
\begin{eqnarray}
(\bm \lambda \cdot {\bm q})^{l} \,
\frac{1}{q^{2}(k^{2}+w^2 q^{2})}=\frac{1}{k^4(w^2+1)}
\sum_{n=0}^{l} C_{l}^{n} ({\mbox{\boldmath $\lambda$}}\cdot{\bm p})^{l-n}
\sum_{m,s=0}^{\infty} \frac{(-1)^{m} (-2)^{s}
({\bm p\cdot\bm k})^{m+s+2} ({\mbox{\boldmath $\lambda$}}\cdot{\bm k})^{n}}
{k^{2(s+m)}}\left(\frac{2w^2}{w^2+1}\right)^m,
\end{eqnarray}
in which we need to collect only the terms proportional to
$({\mbox{\boldmath $\lambda$}}\cdot{\bm p})^{l}$. This leads to the
restriction $n=s+m+2$ and hence to the finite double sum
\begin{eqnarray}
\label{sum1}
(\bm \lambda \cdot {\bm q})^{l} \,
\frac{1}{q^{2}(k^{2}+w^2 q^{2})}&\cong&\frac{1}{k^4(w^2+1)}\\ \nonumber
&\times&\sum_{s,m=0}^{s+m+2\le l} (-1)^{m} (-2)^{s} C_{l}^{s+m+2}
\frac{({\mbox{\boldmath $\lambda$}}\cdot{\bm p})^{l-m-s-2}
({\bm p\cdot\bm k})^{m+s+2} ({\mbox{\boldmath $\lambda$}}\cdot{\bm k})^{s+m+2}}
{k^{2(s+m)}} \left(\frac{2w^2}{w^2+1}\right)^m.
\end{eqnarray}
Substitution of this sum into the expression~(\ref{ioo}) gives rise to the integrals
\begin{eqnarray}
I_{i_{1} \dots i_{2n}} (m) =
\int_{k>m} \frac{d{\bm k}}{(2\pi)^{d}} \, k^{-d-y} \,
\frac{k_{i_{1}} \dots k_{i_{2n}}}{k^{2n}}
\label{Ints}
\end{eqnarray}
with $n=s+m+2 \ge 2$.
They can be found using the expressions
\begin{eqnarray}
I_{i_{1} \dots i_{2n}} (m) = \frac{\delta_{i_{1}i_{2}}\dots
\delta_{i_{2n-1}i_{2n}} + {\rm all\ permutations}}
{d(d+2)\dots (d+2n-2)}\, I(m),
\label{Jnt}
\end{eqnarray}
where $I(m)$ is the scalar integral
\begin{eqnarray}
I(m)= \int_{k>m} \frac{d{\bm k}}{(2\pi)^d}\frac{1}{k^{d+y}} = C_{d}
\frac{m^{-y}}{y}.
\label{ska}
\end{eqnarray}

The sum over all the possible permutations of
$2n$ tensor indices in the numerator of Eq.~(\ref{Jnt}) involves
$(2n-1)!! = (2n)!/2^{n}n!$ terms, but we have to keep only the terms
that give rise to the structure
$({\mbox{\boldmath $\lambda$}}\cdot{\bm p})^{n}$
after the contraction with the vectors {\mbox{\boldmath $\lambda$}} and
${\bm p}$ in the expression~(\ref{sum1}). Hence, there are only
$n!$ such permutations.

Collecting all the factors for the core~(\ref{Hard}) leads to the expression
\begin{eqnarray}
\widetilde{\Delta}\cong -i^{l} ({\mbox{\boldmath $\lambda$}}\cdot{\bm p})^{l}
g \frac{1}{(w^2+1)}C_d\left(\frac{\mu}{m}\right)^{y} \frac{1}{y}
{\cal S}_{l}(d),
\label{FiR}
\end{eqnarray}
where 
\begin{eqnarray}
{\cal S}_{l}(d) = \sum_{s,m=0}^{s+m+2\le l}
\frac{(-1)^{s+m}2^{s} C_{l}^{s+m+2}(s+m+2)!}
{d(d+2)\dots [d+2(s+m)+2]}\left(\frac{2w^2}{w^2+1}\right)^m.
\label{SlApp}
\end{eqnarray}
For $l=0$, sums~(\ref{sum1}) and~(\ref{SlApp}) contain no terms, so that
${\cal S}_{0}(d) = 0$.

Thus, expression~(\ref{Gamma2}) for the functional~(\ref{Gamma1}) reads
\begin{eqnarray}
\Gamma_{2}(x) = F_{l}(x) \left[1 - \frac{g}{2(w^2+1)} \left(\frac{\mu}{m}\right)^{y}\frac{1}{y}\,
{\cal S}_{l}(d)
\right],
\label{Pf}
\end{eqnarray}
where $F_{l}$ is the operator~(\ref{FlSk})
and the substitution $g\to g C_d$ is implied. For the
renormalization constant $Z_l$ in the MS scheme one obtains
\begin{eqnarray}
Z_{l}= 1 - \frac{g}{2(w^2+1)}\frac{1}{y}\, {\cal S}_{l}(d).
\end{eqnarray}
Accordingly to Sec.~\ref{sec:ope} the parameter $l$ defining the composite operator~(\ref{FlSk}) 
counts the Legendre polynomials entering in the OPE for correlation functions.

\subsection{Calculation of the double sum ${\cal S}_{l}(d)$}
\label{b}

It turns out that the double sum ${\cal S}_{l}(d)$ in Eqs.~(\ref{Sl}) and~(\ref{SlApp}) can be reduced
to a simpler
onefold sum. Let us pass from the set of variables $s$ and $m$ to the new
summation variables $k$ and $m$, where
$k=s+m$, and substitute the explicit expression
for the binomial coefficient $C^{k+2}_{l} = l!/(k+2)!(l-k-2)!$. This gives
\begin{eqnarray}
{\cal S}_{l}(d)= l!
\sum_{k=0}^{k+2\le l} \left[ \sum_{m=0}^{k} \left(\frac{w^2}{w^2+1}\right)^m \right]\,
\frac{(-2)^{k}} {(l-k-2)!\,d(d+2)\dots (d+2k+2)}.
\label{sum2}
\end{eqnarray}
The internal summation over $m$ gives
\begin{equation}
\sum_{m=0}^{k}\left(\frac{w^2}{w^2+1}\right)^m = (w^2+1)\left[ 1 - \left(\frac{w^2}{w^2+1}\right)^{k+1}\right];
\end{equation}
changing now the summation variable $k\to k+2$ one obtains
\begin{eqnarray}
{\cal S}_{l}(d)=l!\, (w^2+1)
\sum_{k=2}^{l} \frac{(-2)^{k-2}}{(l-k)!\,d(d+2)\dots (d+2k-2)} \left[ 1 - \left(\frac{w^2}{w^2+1}\right)^{k-1}\right].
\end{eqnarray}
Substitution $z = d/2$ allows us to construct the expression with the ratio of two factorials
\begin{eqnarray}
{\cal S}_{l}(z)=l!\, (w^2+1)\sum_{k=2}^{l} \frac{(-1)^{k}(z-1)!}{4(l-k)!\,(z+k-1)!} \left[ 1 - \left(\frac{w^2}{w^2+1}\right)^{k-1}\right],
\end{eqnarray}
which can be calculated for any given $l$:
\begin{eqnarray}
\label{b5}
{\cal S}_{l}(z)=\frac{l(l-1)}{4z}
\left[\frac{w^2+1}{l+z-1} - \frac{w^2}{z+1}\, {}_2F_1\left(1,2-l;2+z;\frac{w^2}{w^2+1}\right)\right].
\end{eqnarray}
Here ${}_2F_1$ denotes the hypergeomeric function, defined for $|t|<1$ as
\begin{equation}
{}_2F_1(a, b; c; t) = 1 + \sum_{k=1}^{\infty}\left[\prod_{l=0}^{k-1} \frac{(a+l)(b+l)}{(1+l)(c+l)}\right]\, t^k.
\end{equation}
The explicit expression~(\ref{b5}) allows us to analyze the dependence of the anomalous
dimensions $\gamma_l$ over $l$ being in the present context the degree of the anisotropy. 





\section*{References}
\begin{thebibliography}{99}

\bibitem{Legacy} U.~Frisch, {\it Turbulence: The Legacy of A.N.~Kolmogorov} (Cambridge University Press, Cambridge, 1995).

\bibitem{LegacyD} P.~A. Davidson, {\it Turbulence. An introduction for Scientists and Engineers} (Oxford University Press, Oxford, 2005).

\bibitem{Monin} A.~S.~Monin and A.~M.~Yaglom, {\it Statistical fluid mechanics}, vol. 2 (MIT Press, Cambridge, 1975). 

\bibitem{K41a} A.~N. Kolmogorov, Dokl. Akad. Nauk SSSR {\bf 30}(4), 299 (1941); \\
reprinted in Proc. R. Soc. Lond. A {\bf 434}, 9 (1991).

\bibitem{K41b} A.~N. Kolmogorov, Dokl. Akad. Nauk SSSR {\bf 31}(6), 538 (1941).

\bibitem{K41c} A.~N. Kolmogorov, Dokl. Akad. Nauk SSSR {\bf 32}(1), 19 (1941); \\
reprinted in Proc. R. Soc. Lond. A {\bf 434}, 15 (1991).

\bibitem{Richardson} L.~F. Richardson, {\it Weather Prediction by Numerical Process} (Cambridge University Press, Cambridge, 1922).

\bibitem{E84} H.~Effinger and S.~Grossmann, Phys. Rev.~Lett. {\bf 53}, 442 (1984).

\bibitem{S97} K.~R.~Sreenivasan and R.~A.~Antonia, Annu Rev. Fluid. Mech. {\bf 29}, 435 (1997).

\bibitem{P98} I.~Arad, B.~Dhruva, S.~Kurien, V.~L'vov, I.~Procaccia, and K.~R.~Sreenivasan, Phys. Rev.~Lett. {\bf 81}, 5330 (1998).

\bibitem{B96} V.~Borue and S.~A.~Orszag, J.~Fluid Mech. {\bf 306}, 293 (1996).

\bibitem{P99} I.~Arad, V.~L'vov, and I.~Procaccia, Phys. Rev.~E {\bf 59}, 6753 (1999).

\bibitem{P00} I.~Arad, L.~Biferale, and I.~Procaccia, Phys. Rev.~E {\bf 61}, 2654 (2000).

\bibitem{P05} L.~Biferale and I.~Procaccia, Phys. Rep. {\bf 414}, 43 (2005).

\bibitem{FGV} G.~Falkovich, K.~Gaw\c{e}dzki, and M.~Vergassola, Rev. Mod. Phys. {\bf 73}, 913 (2001).

\bibitem{Zinn} J.~Zinn-Justin, {\it Quantum Field Theory and Critical
Phenomena}, 4$^{\rm th}$ ed. (Oxford University Press, Oxford, 2002).

\bibitem{Vasiliev} A.~N.~Vasil'ev, {\it The field theoretic
renormalization group in critical behavior theory and stochastic dynamics}
(Chapman \& Hall/CRC, Boca Raton, 2004).

\bibitem{Tauber} U.~T{\"a}uber, {\it Critical Dynamics: A Field Theory Approach to Equilibrium
      and Non-Equilibrium Scaling Behavior}  (Cambridge University Press, New York, 2014).
      
\bibitem{HHL} M.~Hnati\v{c}, J. Honkonen, T. Lu\v{c}ivjansk\'y, Acta Physica Slovaca {\bf 66}, 69 (2016).

\bibitem{AAV} L.~Ts.~Adzhemyan, N.~V.~Antonov, and A.~N.~Vasil'ev Phys. Rev.~E {\bf 58}, 1823 (1998).

\bibitem{P96} A.~R.~Fairhall, O.~Gat, V.~L'vov, and I.~Procaccia, Phys. Rev.~E {\bf 53}, 3518 (1996).

\bibitem{cube} L.~Ts.~Adzhemyan, N.~V. Antonov, V.~A.~Barinov, Yu.~S.~Kabrits, and A.~N.~Vasil'ev, Phys. Rev.~E {\bf 63}, 025303(R) (2001); \\
Phys. Rev.~E {\bf 64}, 019901(E) (2001); 
Phys. Rev.~E {\bf 64}, 056306 (2001).

\bibitem{FinTime} N.~V.~Antonov, Phys. Rev.~E {\bf 60}, 6691 (1999).

\bibitem{FinTimeEta} L.~Ts.~Adzhemyan, N.~V.~Antonov, J.~Honkonen, Phys. Rev.~E {\bf 66}, 036313 (2002).

\bibitem{AKens} N.~V.~Antonov, Physica~D {\bf 144}, 370 (2000).

\bibitem{JphysA} N.~V.~Antonov, J. Phys. A: Math. Gen. {\bf 39}, 7825 (2006).

\bibitem{Lanotte2-mod2} N.~V.~Antonov and N.~M.~Gulitskiy, Theor. Math. Phys., {\bf 176}(1), 851 (2013).

\bibitem{AGM} N.~V.~Antonov, N.~M.~Gulitskiy, and A.~V.~Malyshev, EPJ Web of Conf. {\bf 126}, 04019 (2016).

\bibitem{R2} M.~Hnatich, J.~Honkonen, M.~Jurcisin, A.~Mazzino, and S.~Sprinc, Phys. Rev.~E {\bf 71}, 066312 (2005).

\bibitem{R2M} E.~Jur\v{c}i\v{s}inova and M.~Jur\v{c}i\v{s}in, Phys. Rev.~E {\bf 77}, 016306 (2008);\\
E.~Jur\v{c}i\v{s}inova, M.~Jur\v{c}i\v{s}in, and R.~Remecky, Phys. Rev.~E {\bf 80}, 046302 (2009).

\bibitem{Arp} H.~Arponen, Phys. Rev.~E {\bf 79}, 056303 (2009).

\bibitem{VectorN}   
N.~V.~Antonov and N.~M.~Gulitskiy,  Phys. Rev.~E {\bf 91},  013002 (2015);
Phys. Rev.~E {\bf 92},  043018 (2015);
AIP Conf. Proc. {\bf 1701}, 100006 (2016); 
EPJ Web of Conf. {\bf 108}, 02008 (2016).

\bibitem{Marian2C}
E.~Jur\v{c}i\v{s}inova and M.~Jur\v{c}i\v{s}in, Phys.~Rev.~E~{\bf 88}, 011004(R) (2013); \\
E.~Jur\v{c}i\v{s}inova, M.~Jur\v{c}i\v{s}in, and M.~Menkyna, Phys. Rev. E {\bf 95}, 053210 (2017).

\bibitem{Tomas} N.~V.~Antonov, N.~M.~Gulitskiy, M.~M.~Kostenko, and T.~Lucivjansky, Phys. Rev.~E {\bf 95}, 033120, (2017);
EPJ Web of Conf. {\bf 125}, 05006 (2016);
EPJ Web of Conf. {\bf 137}, 10003 (2017);
EPJ Web of Conf. {\bf 164}, 07044 (2017).

\bibitem{AK14} N.~V.~Antonov and M.~M.~Kostenko, Phys. Rev.~E {\bf 90}, 063016 (2014).

\bibitem{AK15} N.~V.~Antonov and M.~M.~Kostenko, Phys. Rev.~E {\bf 92}, 053013 (2015). 

\bibitem{HZ} M.~Hnatich and P.~Zalom, Phys. Rev.~E {\bf 94}, 053113 (2016). 

\bibitem{MMZ} E.~Jur\v{c}i\v{s}inova, M.~Jur\v{c}i\v{s}in, and P.~Zalom, Phys. Rev. E {\bf 89}, 043023 (2014).

\bibitem{MMZR} E.~Jur\v{c}i\v{s}inova, M.~Jur\v{c}i\v{s}in, R. Remecky, and P.~Zalom, Phys. Rev. E {\bf 87}, 043010 (2013).

\bibitem{R1} O. G.~ Chkhetiani, M.~Hnatich, E.~Jur\v{c}i\v{s}inova, M.~Jur\v{c}i\v{s}in, A.~Mazzino, and M.~Repa\v{s}an, Phys. Rev.~E {\bf 74}, 036310 (2006).

\bibitem{Marian}
E.~Jur\v{c}i\v{s}inova and M.~Jur\v{c}i\v{s}in, J. Phys. A: Math. Theor., {\bf 45}, 485501 (2012); 
Phys. Part. Nucl. {\bf 44}, 360 (2013);
Phys. Rev.~E {\bf 91}, 063009 (2015); 
Phys. Rev.~E {\bf 94}, 043102 (2016);
Phys. Rev.~E~{\bf 95}, 053112 (2017).

\bibitem{AntGul2012}
N.~V.~Antonov and N.~M.~Gulitskiy, Lecture Notes in Comp. Science, {\bf 7125/2012}, 128 (2012); 
Phys. Rev.~E {\bf 85}, 065301(R)~(2012); 
Phys. Rev.~E {\bf 87}, 039902(E) (2013).

\bibitem{Marian2} E.~Jur\v{c}i\v{s}inova, M.~Jur\v{c}i\v{s}in, and R.~Remeck\'{y}, Phys. Rev. E {\bf 88}, 011002 (2013); Phys. Rev.~E {\bf 93}, 033106 (2016).

\bibitem{AK2} M.~Hnatich, E.~Jur\v{c}i\v{s}inova, M.~Jur\v{c}i\v{s}in,
and M.~Repa\v{s}an, J. Phys. A: Math. Gen. {\bf 39}, 8007 (2006).

\bibitem{turbo} L.~Ts.~Adzhemyan, N.~V.~Antonov, A.~N.~Vasil'ev, \emph{The Field Theoretic Renormalization Group in Fully Developed
Turbulence} (Gordon \& Breach, London, 1999).

\bibitem{OU1} N.~G. Van Kampen, \emph{Stochastic Processes in Physics and Chemistry}, 3$^{\rm rd}$ ed. (North Holland, 2007).

\bibitem{OU2} C. Gardiner, \emph{Stochastic Methods: A Handbook for the Natural and Social Sciences}, 4$^{\rm th}$ ed. (Springer-Verlag Berlin Heidelberg, 2009).

\bibitem{NonG} M.~Holzer and E.~D.~Siggia, Phys. Fluids {\bf 6}, 1820 (1994).

\bibitem{Carati} D.~Carati, Phys. Fluids~A {\bf 2}, 1854 (1990).

\bibitem{CaratiV} S. Gama and M. Vergassola, Physica~D {\bf 76}, 291 (1994).

\bibitem{CaratiK} E.~Medina, T.~Hwa, M.~Kardar, and Y.-C.~Zhang, Phys. Rev.~A {\bf 39}, 3053 (1989).

\bibitem{Astro1} J.~F.~Barbero~G., A.~Dominguez, T.~Goldman, and J.~Perez-Mercader, Europhys. Let. {\bf 38}(8), 637 (1997).

\bibitem{Astro2} A.~Dominguez, D.~Hochberg, J.~M.~Martin-Garcia, J.~Perez-Mercader, and L.~S.~Schulman, Astron. Astrophys. {\bf 344}, 27~(1999); 
Astron. Astrophys. {\bf 363}, 373 (2000).

\bibitem{Kim} L.~Ts.~Adzhemyan, N.~V.~Antonov, J.~Honkonen, and T.~L.~Kim, Phys. Rev.~E {\bf 71}, 016303 (2005).

\bibitem{LM} A.~Lanotte and A.~Mazzino, Phys. Rev.~E {\bf 60}, R3483(R) (1999).

\bibitem{ALM} N.~V.~Antonov, A.~Lanotte, and A.~Mazzino, Phys. Rev.~E {\bf 61}, 6586 (2000).

\bibitem{AHMMG} N.~V.~Antonov, J.~Honkonen, A.~Mazzino, and P.~Muratore-Ginanneschi, Phys. Rev.~E {\bf 62}, R5891(R) (2000).

\bibitem{Two} D. Ronis, Phys. Rev. A {\bf 36}, 3322 (1987); \\
J. Honkonen and M. Yu. Nalimov, {Z. Phys.} B {\bf 99}, 297 (1996); \\
L.~Ts. Adzhemyan, J. Honkonen, M.~V. Kompaniets, and A.~N. Vasil'ev, Phys. Rev. E {\bf 71}, 036305 (2005).

\bibitem{NS-Zero} L.~Ts.~Adzhemyan, A.~N.~Vasil'ev, Yu.~M.~Pis'mak, Theor. Math. Phys., {\bf 57}(2), 1131 (1983).

\bibitem{InfD1} J.-D. Fournier, U. Frisch, and H. A. Rose, J. Phys. A: Math. Gen. {\bf 11}, 187 (1978).

\bibitem{InfD2} H.~L. Frisch and M. Schultz, Physica A {\bf 211}, 37 (1994). 

\bibitem{InfD3} L.~Ts.~Adzhemyan, N.~V.~Antonov, P.~B.~Gol'din, T.~L.~Kim, and M.~V.~Kompaniets, J. Phys. A: Math. Theor. {\bf 41}, 495002 (2008).

\bibitem{InfD4} L.~Ts.~Adzhemyan, N.~V.~Antonov, P.~B.~Gol'din, and M.~V.~Kompaniets, J. Phys. A: Math. Theor. {\bf 46}, 135002 (2013).

\bibitem{AVH84} L.~Ts.~Adzhemyan, A.~N.~Vasil'ev, M.~Hnatich, Theor. Math. Phys. {\bf 58}(1), 47 (1984).

\bibitem{Jackson} J.~D.~Jackson, {\it Classical Electrodynamics}, 3$^{\rm rd}$ ed. (John Wiley and Sons, New York, 1998).


\end{thebibliography}
\end{document}